\documentclass[reprint,superscriptaddress, amsmath,amssymb,aps]{revtex4-2}

\usepackage{graphicx}
\usepackage{dcolumn}
\usepackage{bm}
\usepackage{gensymb}
\usepackage{braket}
\usepackage{xcolor}
\usepackage[colorlinks=true,citecolor=blue,linkcolor=blue,urlcolor=blue]{hyperref}

\begin{document}

\title{Machine-Learning Spin-Fermion Models for Strongly Disordered Magnets: Sb-Doped MnTe}

\author{Do Hoon Kiem}
\email{kimd4@ornl.gov}
\affiliation{Materials Science and Technology Division, Oak Ridge National Laboratory, Oak Ridge, Tennessee 37831, USA}

\author{Parul R. Raghuvanshi}
\affiliation{Materials Science and Technology Division, Oak Ridge National Laboratory, Oak Ridge, Tennessee 37831, USA}

\author{David S. Parker}
\affiliation{Materials Science and Technology Division, Oak Ridge National Laboratory, Oak Ridge, Tennessee 37831, USA}

\author{Lucas Lindsay}
\affiliation{Materials Science and Technology Division, Oak Ridge National Laboratory, Oak Ridge, Tennessee 37831, USA}

\author{Valentino R. Cooper}
\affiliation{Materials Science and Technology Division, Oak Ridge National Laboratory, Oak Ridge, Tennessee 37831, USA}

\author{Tom Berlijn}
\email{berlijnt@ornl.gov}
\affiliation{Center for Nanophase Materials Sciences, Oak Ridge National Laboratory, Oak Ridge, Tennessee 37831, USA}

\date{\today}

\begin{abstract}
Metallic altermagnets are promising for spintronics, but most altermagnets are insulating. Doping can induce metallicity, but introduces chemical and magnetic disorder that is challenging to model from first principles. Here, we introduce a machine-learning approach that renormalizes first-principles calculations onto disordered spin-fermion models. The learned models accurately reproduce first-principles magnetic and electronic properties and enable predictions of millions of inequivalent long-range exchange couplings in large disordered supercells. Applied to Sb-doped MnTe, the approach shows that altermagnetism persists up to about 25\% Sb, and reveals disorder broadening as a new factor controlling spin selectivity. More broadly, our work opens a route to combining renormalization with machine-learned Hamiltonians for strongly disordered materials.
\end{abstract}


\maketitle

\section{\label{sec:introduction}Introduction}

\begin{figure}[!htbp]
    \centering
    \includegraphics[width=1.0\linewidth]{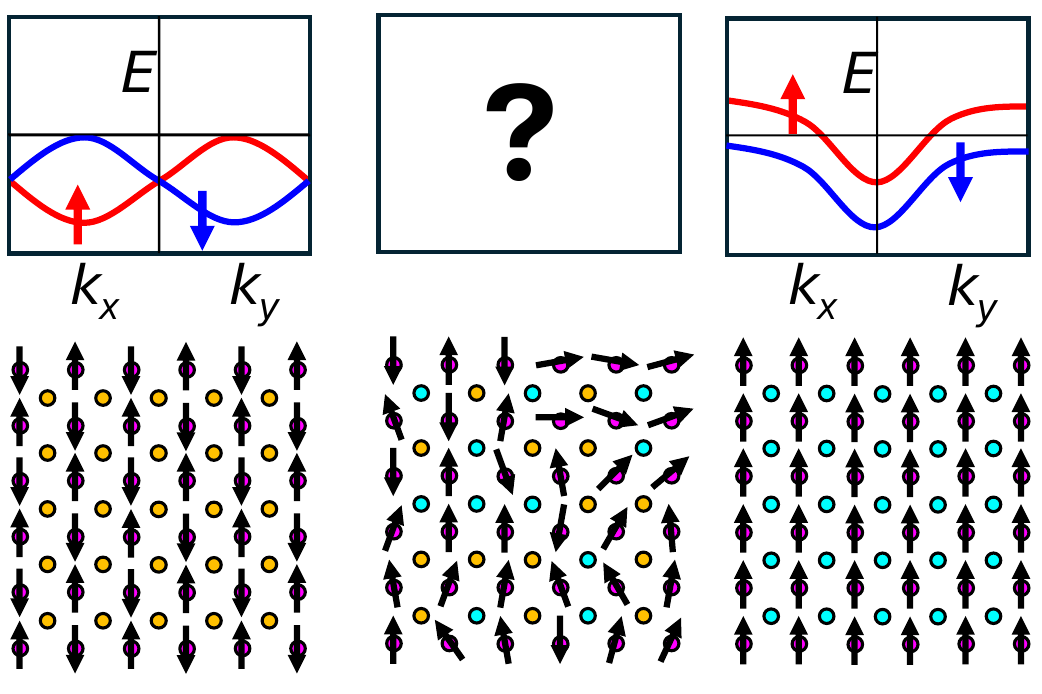}
    \caption{Schematic illustration of interplay between altermagnetism and metallicity in Sb-doped MnTe. MnTe (left) is an insulating altermagnet, while MnSb (right) favors metallic ferromagnetism; intermediate doping introduces chemical and magnetic disorder difficult to capture with conventional simulations.
}
    \label{fig1:concept}
\end{figure}

Altermagnets (AMs)~\cite{Naka2019,Hayami2019,Hayami2020,Smejkal2022beyond,Smejkal2022landscape} are promising for nonvolatile magnetic random-access memory (MRAM), combining the strong spin-transfer torques of ferromagnets (FMs), which enable efficient writing, with the negligible stray fields of antiferromagnets (AFMs), which allow denser bit packing~\cite{GonzalezHernandez2021,Smejkal2022GMR,Bai2022SST,Karube2022}. Altermagnets may further enable unconventional superconductivity, including spin-triplet pairing, which can host topological superconductivity and Majorana modes of interest for fault-tolerant quantum computing~\cite{Zhu2023TSC,Brekke2023,Ghorashi2024}. Both applications require metallicity: MRAM relies on electrical currents for reading and writing, while superconductivity requires mobile charge carriers. However, most predicted altermagnets are insulating~\cite{Wan2025metallicAM,Sufyan2026}, limiting their potential for these applications.

An alternative strategy for realizing metallic altermagnets is to dope insulating altermagnets, but the effects of doping on magnetic order can be difficult to predict. For example, Sb doping drives MnTe, an insulating altermagnet~\cite{Mu2019MnTe,Krempasky2024,Lee2024MnTe}, toward MnSb, a metallic ferromagnet~\cite{Continenza2001,Taylor2015MnSb,Marenkin2018}. How does altermagnetism evolve between these two limits (Fig.~\ref{fig1:concept})? Answering this question at the microscopic level requires first-principles simulations that capture the complex interplay between chemical disorder and long-range magnetic interactions. In metallic systems, these interactions are mediated by conduction electrons (Ruderman--Kittel--Kasuya--Yosida, or RKKY, interactions~\cite{Ruderman1954,Kasuya1956,Yosida1957}) and extend across many inequivalent chemical environments, giving rise to massive numbers of distinct magnetic couplings and complex noncollinear ground states. Furthermore, modeling chemical and magnetic disorder requires large supercells, which fold electronic bands into smaller Brillouin zones and obscure the momentum-dependent spin splitting characteristic of altermagnets. Band unfolding can recover this momentum-resolved information~\cite{Ku2010unfolding,Popescu2010}, but accurately resolving disorder-induced spectral broadening requires calculations for many large supercells. Our recent first-principles calculations of Sb-doped MnTe illustrate these limitations: we found an antiferromagnetic-to-ferromagnetic crossover near 25\% Sb, but were restricted to a single 128-atom supercell per composition and to collinear magnetic configurations~\cite{Raghuvanshi2026}. Together, the complex long-range magnetic interactions, noncollinear magnetic ground states, and need for extensive supercell calculations make conventional first-principles approaches computationally prohibitive.

Machine learning offers a way forward. Graph neural networks (GNNs) based on message passing~\cite{Gilmer2017} are now widely used to learn total energies and forces ~\cite{Behler2007,Zhang2018DeePMD,Batzner2022}, and have since been extended to electronic Hamiltonians~\cite{Schutt2019SchNOrb,Wang2021TBML,Li2022DeepH}. These machine-learned Hamiltonian approaches have followed two main routes. 
The first learns the DFT Hamiltonian matrix directly, as in SchNOrb~\cite{Schutt2019SchNOrb}, DeepH~\cite{Li2022DeepH} and its E(3)-equivariant successors such as DeepH-E3, HamGNN, QHNet and MACE-H~\cite{Gong2023DeepHE3,Zhong2023HamGNN,Yu2023QHNet,Qian2026MACEH}. These methods bypass self-consistency but retain the atomic-orbital basis of the underlying DFT calculation, so the Hilbert space is not reduced. The second approach reduces the Hilbert space by fitting minimal-basis tight-binding models to band energies or other electronic observables~\cite{Wang2021TBML,Schattauer2022,Gu2024DeePTB,Cignoni2024,Choudhary2025SlaKoNet}. These methods learn the reduced Hamiltonian only indirectly, through the observables it should reproduce, rather than training directly on renormalized Hamiltonians, and they are restricted to electronic structure and cannot be extended to other low-energy degrees of freedom, such as spins.  A key advance for magnetic systems is xDeepH~\cite{Li2023xDeepH}, together with the related HamGNN++~\cite{Zhong2023HamGNNpp}, which learn the DFT Hamiltonian as a function of both atomic positions and magnetic moments, but they require the magnetic configuration as input, which is not known in advance in disordered systems. 
This configuration could be obtained from energy-based machine-learning models that also depend on the spin directions~\cite{Yu2024SpinGNN,Yuan2024MagNet,Yang2023DeePSPIN}. These models capture the magnetic interactions through total energies and magnetic torques computed for many constrained spin configurations of each structure, which becomes demanding in metals, where long-range RKKY-type interactions couple many spins. An alternative is to bring renormalization, the reduction of many degrees of freedom to a few that capture the essential physics, into machine-learned Hamiltonian approaches, so that the magnetic interactions are learned explicitly.

Here, we introduce a machine-learning framework that renormalizes first-principles calculations of strongly disordered magnets onto disordered spin-fermion models, and apply it to Sb-doped MnTe. Trained on small cells, the spin model predicts millions of long-range exchange couplings in large disordered supercells, and the electronic model predicts the corresponding electronic Hamiltonians. Benchmarks against first-principles calculations confirm the accuracy of both. We track the evolution of altermagnetism with doping by applying the framework to many large supercells across a range of Sb concentrations and unfolding the resulting electronic structure into the primitive Brillouin zone. We find that altermagnetism persists up to 25\% Sb doping, but disorder broadens the electronic bands such that the tail of each spin channel extends into the peak of the other. This adds a new dimension to altermagnetism: spin selectivity is controlled not only by the spin splitting but also by disorder broadening, which must therefore be considered in the design of doped altermagnets for applications. More broadly, our results demonstrate that combining renormalization with machine-learned Hamiltonians is a viable route to constructing effective models of structurally complex materials beyond the reach of direct first-principles calculations.

\section{\label{sec:results} Results}

\subsection{Machine-learning disordered spin-fermion models}

\begin{figure}[!htbp]
    \centering
    \includegraphics[width=1.0\linewidth]{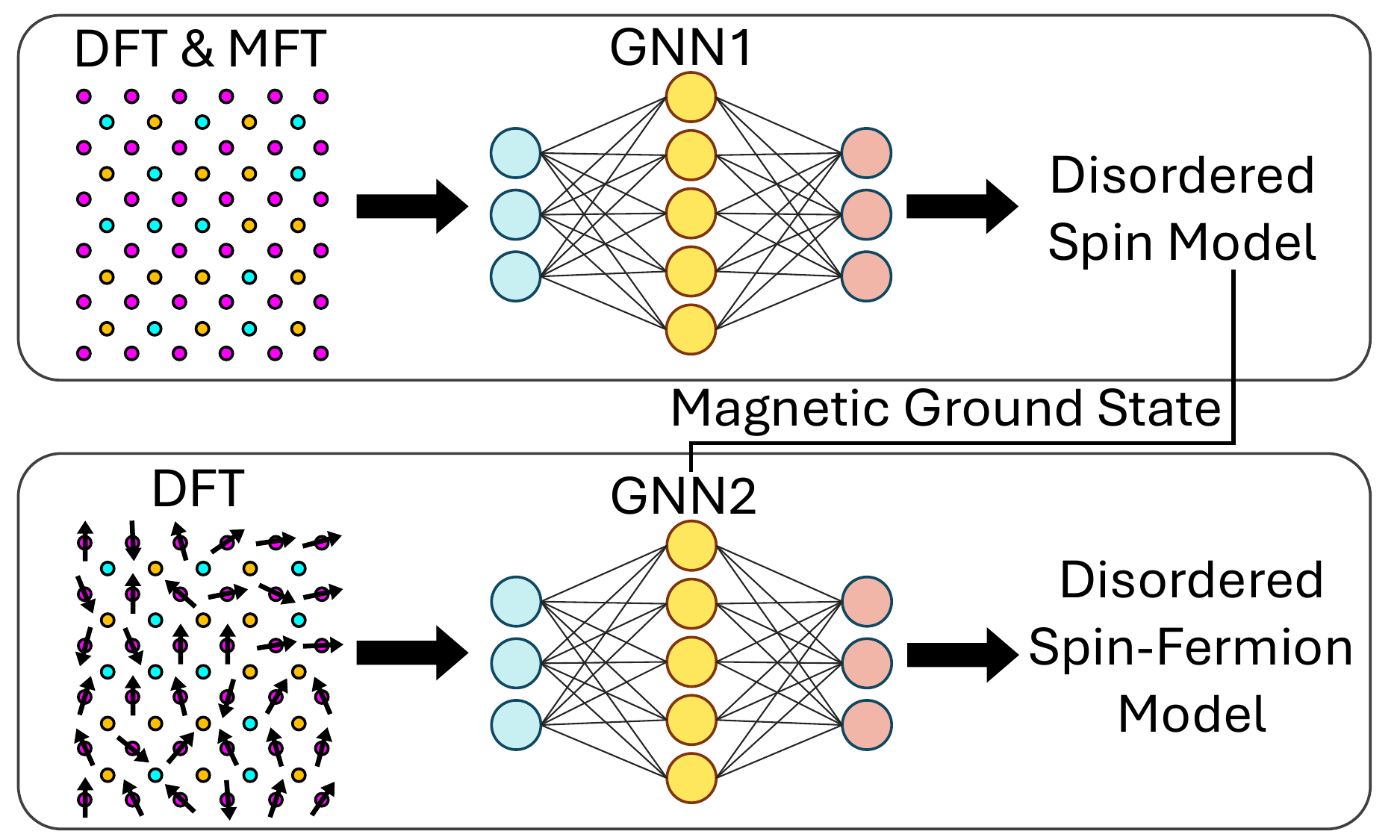}
    \caption{Machine-learning spin-fermion models for strongly disordered magnets. Upper row: DFT and the magnetic force theorem (MFT) provide exchange couplings used to train GNN1, which predicts spin Hamiltonians of large disordered supercells. Lower row: DFT provides electronic Hamiltonians used to train GNN2, which predicts electronic Hamiltonians of large disordered supercells. The magnetic ground state obtained from the spin Hamiltonian is used as input to GNN2. Together, the two networks define a disordered spin-fermion model.}
    \label{fig2:GNN}
\end{figure}

To map first-principles calculations of strongly disordered magnets onto spin-fermion models, we use two graph neural networks (Fig.~\ref{fig2:GNN}). It is trained on DFT calculations of MnTe$_{1-x}$Sb$_x$ supercells with varying cell lengths and angles, Sb concentrations, and Sb arrangements. For each supercell, the exchange couplings $J_{ij}$ are extracted with the magnetic force theorem (MFT)~\cite{Liechtenstein1987,Katsnelson2000,Szilva2023}. Given only the atomic positions, GNN1 then predicts all exchange couplings of a supercell, defining a disordered spin model. In this work we restrict this model to isotropic Heisenberg exchange, $H_{\mathrm{spin}} = -\sum_{i<j} J_{ij}\,\mathbf{S}_i\cdot\mathbf{S}_j$, although anisotropic and single-ion terms can be included in the same framework. The second network, GNN2, learns the electronic model. It is trained on DFT Hamiltonians computed for supercells with both chemical disorder and disordered spin configurations, and predicts the electronic Hamiltonian as a function of both the atomic positions and the orientations of the magnetic moments.

The two networks are combined as follows. For a given disordered supercell, GNN1 provides the spin model, whose magnetic ground state is then determined by torque relaxation of the classical spins using damped Landau--Lifshitz--Gilbert (LLG) dynamics~\cite{Gilbert2004}. Because this step involves only classical spins, it is computationally affordable even for large supercells with many inequivalent couplings, and in general yields complex noncollinear ground states. This ground state is passed to GNN2, which predicts the electronic Hamiltonian in the presence of both chemical disorder from Sb substitution and magnetic disorder from the resulting spin texture. Together, the two networks map DFT onto a disordered spin-fermion model with classical spins, in which the electrons move in the magnetic ground state determined by the spin model. We do not relax the atomic structures in this work: relaxing each training configuration would substantially increase the cost of generating thousands of training configurations used for the two networks, and the additional structural variability would likely require even more of them. Instead, Sb atoms occupy the ideal lattice sites of MnTe, with the experimental lattice constants $a = 4.15$~\AA\ and $c = 6.71$~\AA~\cite{Szuszkiewicz2006}. Because both networks take atomic positions as input, relaxed structures can be included directly in future work. Details of the DFT calculations, network training and ground-state relaxations are given in Methods.

\subsection{Spin model: benchmarking and magnetic phase evolution}

\begin{figure}[!htbp]
    \centering
    \includegraphics[width=1.0\linewidth]{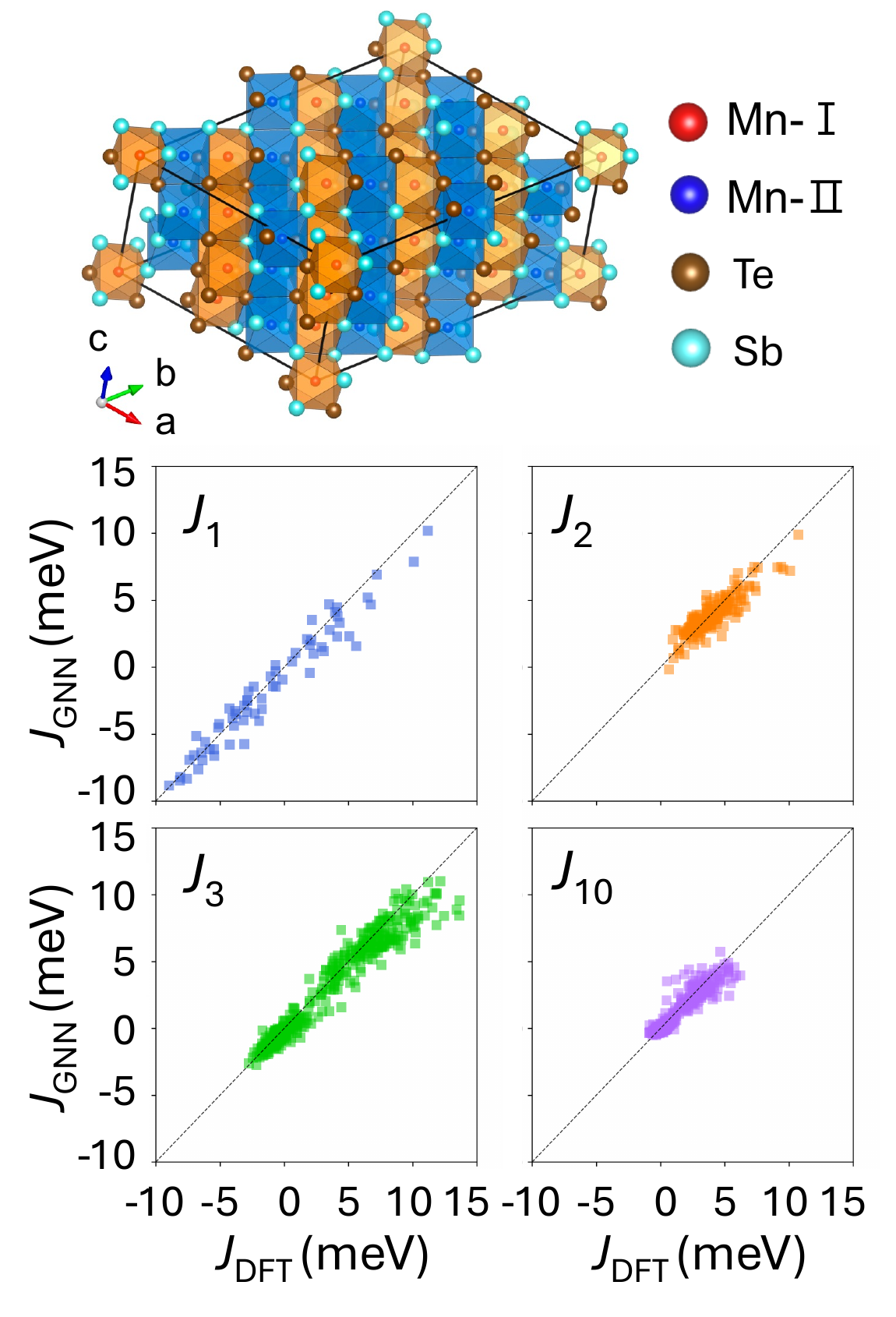}
		\caption{Benchmark of GNN predictions of Heisenberg exchange interactions. Top: Non-orthogonal Mn$_{64}$Te$_{32}$Sb$_{32}$ test supercell. Mn-I and Mn-II (red and blue spheres, with orange and blue octahedra, respectively) denote the Mn sublattices with spin up and spin down in the altermagnetic state. The axes $a$, $b$ and $c$ indicate the supercell lattice vectors; the $c$ axis of the MnTe unit cell (not shown) is horizontal. Bottom: GNN-predicted versus DFT exchange interactions, $J_{\mathrm{GNN}}$ versus $J_{\mathrm{DFT}}$, for the first-, second-, third-, and tenth-nearest-neighbor couplings $J_1$, $J_2$, $J_3$ and $J_{10}$. The dashed line indicates perfect agreement.}		
    \label{fig3:benchmarkJ}
\end{figure}

Before turning to the GNN, we compare the DFT exchange couplings of the end member MnTe with inelastic neutron scattering (Supplementary Note~S5~\cite{SI}). The dominant couplings $J_1$ and $J_3$ agree well with experiment~\cite{Szuszkiewicz2006,Liu2024MnTe}, while the weak in-plane coupling $J_2$ is overestimated, a coupling that previous DFT studies also fail to reproduce~\cite{Mu2019MnTe,Mazin2023MnTe,Alaei2024MnTe}. For the end member MnSb, a direct comparison is less meaningful: neutron scattering constrains only linear combinations of the couplings, the samples contain interstitial Mn, and the analysis neglects induced moments on Sb~\cite{Yamaguchi1992MnSb,Taylor2015MnSb}. Our couplings nevertheless reproduce its strongly ferromagnetic character. Because our aims are to show that machine-learned spin-fermion models can reproduce DFT, and to capture the qualitative evolution from altermagnetic MnTe to ferromagnetic MnSb, this level of agreement with experiment is sufficient for the present purpose.

Turning to the GNN, we computed exchange couplings up to the tenth-nearest neighbor for 510 MnTe$_{1-x}$Sb$_x$ supercells with Sb concentrations from 2 to 50\% (${\sim}80$ atoms on average) to train GNN1. Here $J_n$ denotes the coupling between Mn atoms in the $n$th neighbor shell, ordered by Mn--Mn distance, with $J_1$ the interlayer coupling along $c$. The supercells are non-orthogonal and vary in lattice constants and angles, so that the training data are not biased toward any particular supercell geometry. Figure~\ref{fig3:benchmarkJ} compares GNN and DFT exchange couplings for a Mn$_{64}$Te$_{32}$Sb$_{32}$ test supercell not included in the training set, for $J_1$, $J_2$, $J_3$ and $J_{10}$ (for $J_4$ to $J_9$, see Supplementary Fig.~S1~\cite{SI}). The root-mean-square errors (RMSEs) for $J_1$, $J_2$, $J_3$ and $J_{10}$ in this supercell are 1.20, 0.81, 1.08 and 0.59~meV, only 6--8\% of the ranges of 20, 10, 16 and 7~meV spanned by these couplings. The benchmark also illustrates the complexity of the disordered spin model. Instead of ten exchange couplings, as in pristine MnTe, this single supercell contains 2624 distinct couplings, whose magnitudes vary widely and whose signs can even reverse, as seen for $J_1$. Moreover, the couplings are long ranged. While $J_4$ to $J_9$ are small, $J_{10}$, between Mn atoms 9.8~\AA\ apart, reaches up to 5~meV, comparable to $J_1$, consistent with RKKY-type exchange. Couplings beyond $J_{10}$ are again small (Supplementary Fig.~S2~\cite{SI}), so we truncate the spin model at the tenth neighbor to limit the computational cost of training.

\begin{figure}[!htbp]
    \centering
    \includegraphics[width=1.0\linewidth]{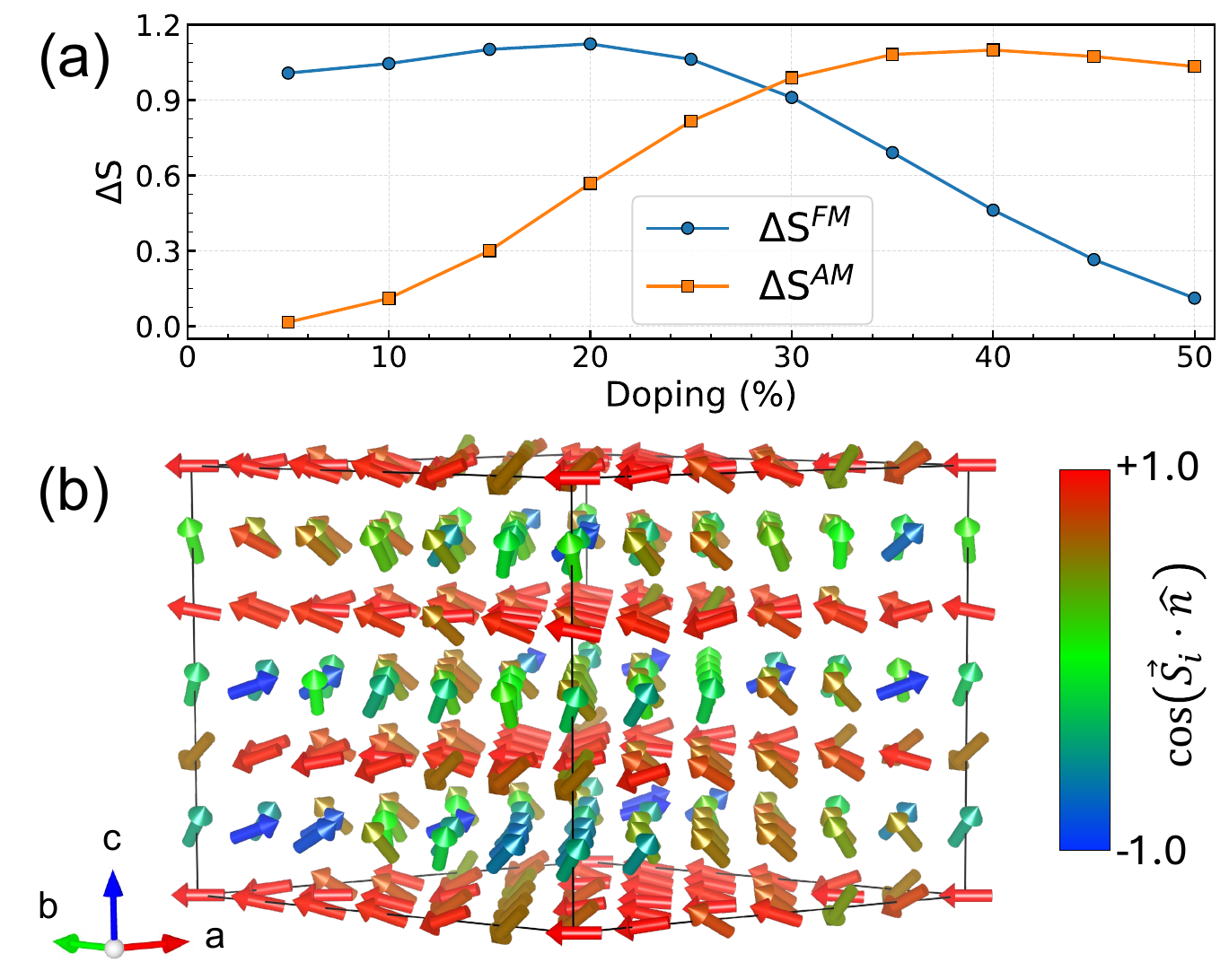}
    \caption{Magnetic phase evolution in Sb-doped MnTe. (a) Deviation of the magnetic ground state from altermagnetic (AM) and ferromagnetic (FM) reference states, $\Delta S^\mathrm{Ref} = \frac{1}{N}\sum_i |\mathbf{S}_i^\mathrm{GS}-\mathbf{S}_i^\mathrm{Ref}|$ (Ref = AM or FM), averaged over spin sites and 50 disorder configurations of $6\times6\times4$ supercells. The crossing of $\Delta S^\mathrm{AM}$ and $\Delta S^\mathrm{FM}$ at roughly 25--30\% Sb marks the AM-to-FM crossover. (b) Magnetic ground state of a representative $6\times6\times4$ supercell at 30\% Sb. Colors show the cosine of the angle between each spin and $\hat{n}\parallel \hat{b}-\hat{a}$.}
    \label{fig4:phasediagram}
\end{figure}

We next use the predicted exchange couplings to determine the magnetic ground states of large disordered supercells. Due to the fact that the spin model is cheap to solve, we can treat $6\times6\times4$ supercells (576 atoms) and average over 50 disorder configurations for each Sb concentration. Counting all couplings up to the tenth-nearest neighbor (82 per primitive cell), each $6\times6\times4$ supercell contains 11,808 inequivalent exchange couplings, amounting to about 6 million over the 50 configurations and 10 Sb concentrations considered. Obtaining the same ground states directly from DFT would be prohibitive in practice, because it would require noncollinear self-consistent calculations for many large supercells, which are both expensive and notoriously difficult to converge. To track the evolution of the magnetic order, we compare each ground state with altermagnetic (AM) and ferromagnetic (FM) reference states using $\Delta S^{\mathrm{Ref}} = \frac{1}{N}\sum_i |\mathbf{S}_i^{\mathrm{GS}} - \mathbf{S}_i^{\mathrm{Ref}}|$, averaged over spin sites and disorder configurations (Fig.~\ref{fig4:phasediagram}a). At 0\% Sb the ground state is the AM state of pristine MnTe, while at 50\% Sb it is almost fully FM. Between these limits, the two measures cross at roughly 25--30\% Sb, marking the AM-to-FM crossover, consistent with our earlier collinear DFT calculations~\cite{Raghuvanshi2026}. Beyond locating the crossover, however, the spin model reveals its nature: Figure~\ref{fig4:phasediagram}b shows a representative ground state at 30\% Sb. Instead of the antiparallel stacking of AM layers or the parallel stacking of FM layers, the spins in consecutive Mn layers are rotated by roughly 90$^\circ$ with respect to each other, with additional local canting in all directions. Close to the crossover, the disordered ground state is therefore a complex noncollinear state intermediate between AM and FM order.

\subsection{Spin-fermion model: benchmarking and electronic-structure evolution}

\begin{figure}[!htbp]
    \centering
    \includegraphics[width=1.0\linewidth]{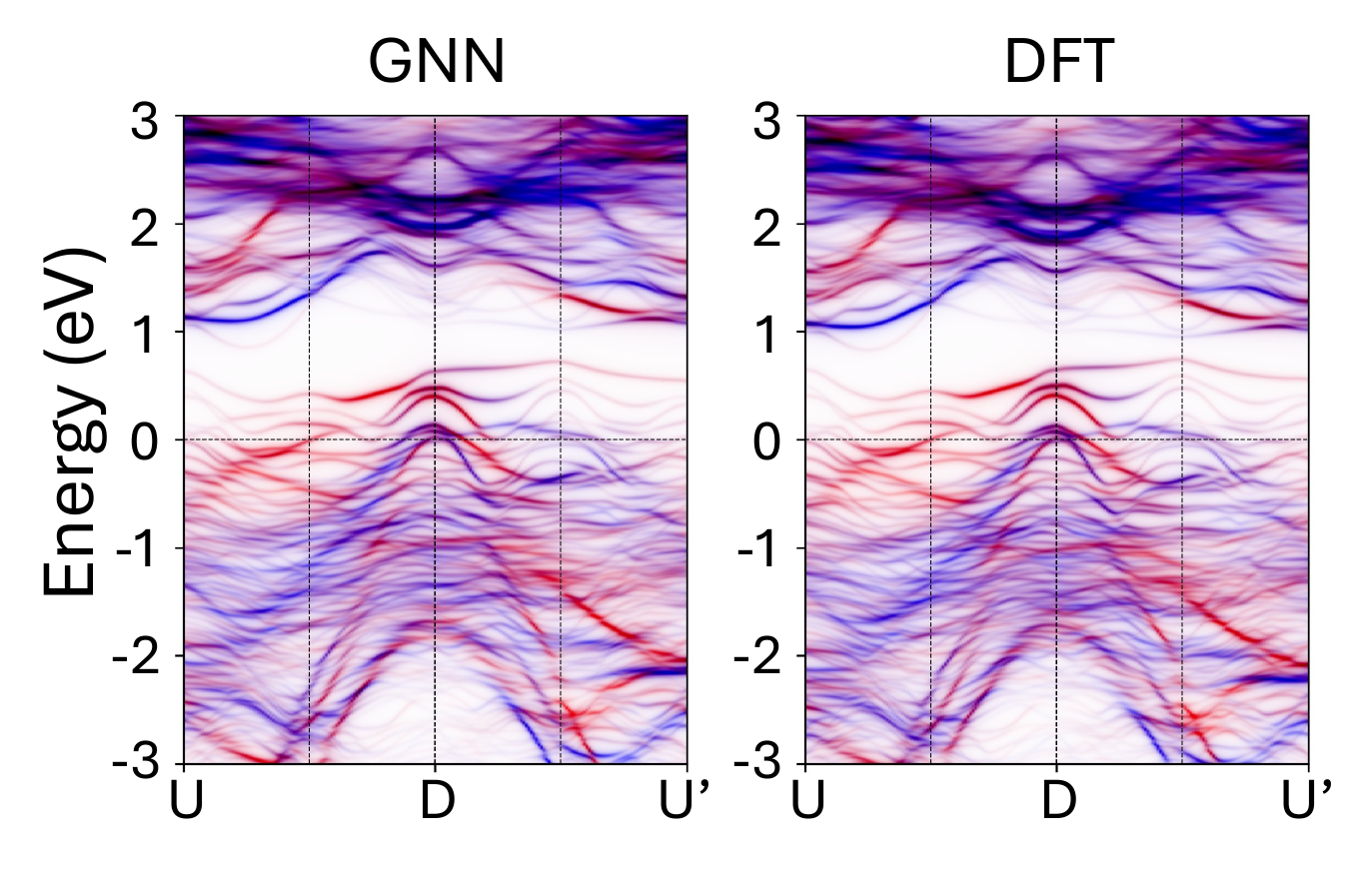}
    		\caption{Benchmark of the electronic GNN. Unfolded spectral functions $A^{\sigma}(\mathbf{k},E)$ of a Mn$_{16}$Te$_{12}$Sb$_{4}$ test supercell (25\% Sb) with disordered spin configuration, computed with the GNN (left) and DFT (right). Red and blue denote $\sigma=\uparrow$ and $\downarrow$, respectively.}
    \label{fig5:benchmarkelec}
\end{figure}

GNN2 was trained on 2040 non-orthogonal supercells (${\sim}80$ atoms on average) with Sb concentrations from 2 to 50\% and disordered spin configurations. As a benchmark, Fig.~\ref{fig5:benchmarkelec} compares the unfolded band structure from GNN2 with that from DFT for a Mn$_{16}$Te$_{12}$Sb$_{4}$ supercell (25\% Sb) not included in the training set. The spectral functions are shown along the path $U$--$D$--$U'$, with $D = (0, 0, 0.3)$, $U = (0.5, 0, 0.3)$ and $U' = (0, 0.5, 0.3)$ in units of the reciprocal lattice vectors (Supplementary Fig.~S3~\cite{SI}). This path is analogous to that used in the angle-resolved photoemission study of Ref.~\cite{Lee2024MnTe}, which lies at $k_z = 0.15$ in the same units. Both lie away from the spin-degenerate nodal planes at $k_z = 0$ and $0.5$, so that they show the altermagnetic spin splitting. The GNN reproduces all the complex features of the DFT spectral function, including the gap openings and shadow bands induced by chemical and magnetic disorder. Because this test supercell contains only 32 atoms, its short periodicity produces pronounced band reconstructions with large gap openings. In larger supercells these features merge into smoothly broadened spectral functions, as discussed below. The agreement is weaker at 50\% Sb (Supplementary Fig.~S4~\cite{SI}). There, the GNN still reproduces the pattern of gap openings and shadow bands, but the band energies are shifted relative to DFT. This is understandable, because 50\% Sb lies at the edge of the training set, and the accuracy could presumably be improved by including more training data at and beyond 50\%. Since this work focuses on the altermagnetic-to-ferromagnetic crossover near 25\%, where the agreement is good, we leave such improvements for future work.

\begin{figure*}[!t]
    \centering
    \includegraphics[width=1.0\linewidth]{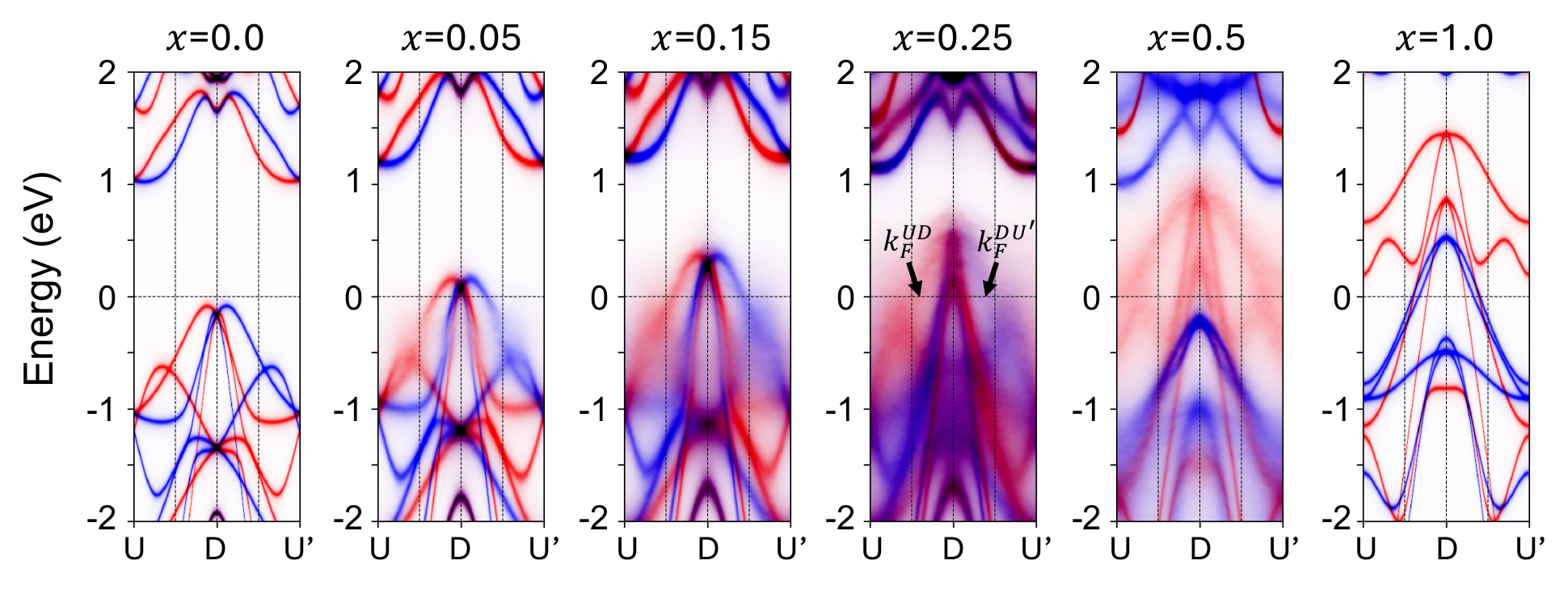}
		\caption{Evolution of the electronic structure from altermagnetic MnTe to ferromagnetic MnSb. Configuration-averaged unfolded spectral functions $\langle A^{\sigma}(\mathbf{k},E)\rangle$ of MnTe$_{1-x}$Sb$_x$ for $x = 0$, 0.05, 0.15, 0.25, 0.5 and 1. Spectra for intermediate concentrations are obtained from the spin-fermion model and averaged over 100 non-orthogonal supercells with ${\sim}120$ atoms on average. Red and blue denote $\sigma=\uparrow$ and $\downarrow$, respectively. Arrows in the $x = 0.25$ panel mark the outermost Fermi momenta $k_F^{UD}$ and $k_F^{DU'}$, located 60\% of the way from $U$ to $D$ and from $U'$ to $D$, respectively, analyzed in Fig.~\ref{fig8:spinmix}.}
		    \label{fig6:spectral}
\end{figure*}

With the spin-fermion model as a surrogate for DFT, we can now study how the electronic structure evolves from MnTe to MnSb. Figure~\ref{fig6:spectral} shows the unfolded spectral functions of MnTe$_{1-x}$Sb$_x$ from $x = 0$ to $x = 1$, where $x$ is the Sb fraction (e.g., $x = 0.25$ corresponds to 25\% Sb). For each intermediate Sb concentration, the magnetic ground state is obtained from GNN1 and the electronic Hamiltonian from GNN2, and the spectral function is averaged over 100 non-orthogonal supercells with ${\sim}120$ atoms on average. This averaging over many large supercells converges the gap openings of individual supercells into smoothly broadened spectral functions. Varying the supercell shape is important here: for supercells of a single shape, gaps open at the same supercell Brillouin-zone boundaries regardless of how the impurities are distributed within the supercell~\cite{Berlijn2011NaxCoO2}.

Starting with undoped MnTe on the left of Fig.~\ref{fig6:spectral}, the spectral function shows the sharp, spin-split bands of a pure altermagnet. As the Sb concentration increases, the bands broaden, reflecting the finite lifetimes and mean free paths of carriers scattered by chemical disorder from Sb substitution and by magnetic disorder in the ground state. Remarkably, altermagnetism persists up to 25\% Sb doping: even there, the momentum-resolved spin selectivity of the altermagnetic bands remains visible, with predominantly spin-up spectral weight near the midpoint of the $U$--$D$ path and predominantly spin-down weight near the midpoint of the $D$--$U'$ path. At 50\% Sb, consistent with Fig.~\ref{fig4:phasediagram}, the altermagnetic order has given way to ferromagnetism: the bands crossing the Fermi level carry spin-up weight along both the $U$--$D$ and $D$--$U'$ paths. The 50\% spectral function already resembles that of MnSb, whose sharp, unbroadened ferromagnetic bands are shown at 100\%. We can now answer the question raised in the introduction: how does altermagnetism evolve as Sb is introduced into MnTe? We find that altermagnetism extends up to 25\% Sb doping, albeit with significant disorder-induced broadening, whose influence we discuss next.

\begin{figure}[!htbp]
    \centering
    \includegraphics[width=1.0\linewidth]{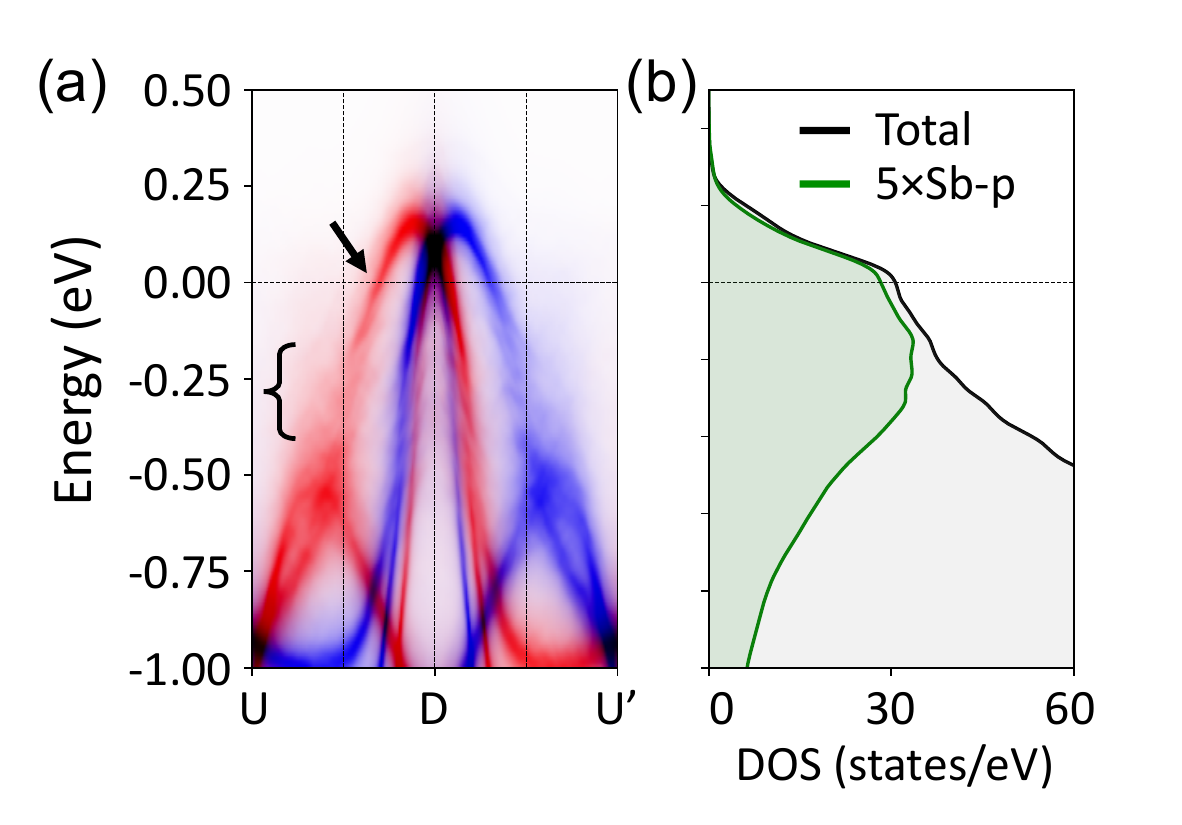}
\caption{Energy-selective disorder broadening at 5\% Sb. (a) Zoom-in of the configuration-averaged unfolded spectral function $\langle A^{\sigma}(\mathbf{k},E)\rangle$ of Fig.~\ref{fig6:spectral}. Bands near the midpoint of the $U$--$D$ path at about $-300$~meV are strongly broadened (curly bracket), whereas the bands crossing the Fermi level remain sharp (black arrow). (b) Total and Sb-$p$ partial density of states; the Sb-$p$ contribution is multiplied by five for clarity.}
    \label{fig7:sharp}
\end{figure}

Focusing first on 5\% Sb (Fig.~\ref{fig7:sharp}) we notice a remarkable feature. Near the midpoint of the $U$--$D$ path, while the bands around $-300$~meV are strongly broadened (curly bracket in Fig.~\ref{fig7:sharp}a), the bands crossing the Fermi level remain sharp (black arrow), and the bands below about $-500$~meV are sharp again. A possible explanation could be that the strongly broadened bands have a large Sb-$p$ contribution. To test this, Fig.~\ref{fig7:sharp}b compares the Sb-$p$ partial density of states with the total density of states. The Sb-$p$ weight does peak near $-200$~meV, close to the strongly broadened bands, but the correlation is not precise: the Sb-$p$ weight is nearly as large at the Fermi level, where the bands remain sharp. The Sb-$p$ contribution alone therefore cannot explain this energy-dependent broadening. Other factors, such as the phase space available for scattering or the orbital character of the host bands, may play a role, and a detailed understanding requires further study. While our simulations do not yet include atomic relaxations, they reveal that disorder broadening in doped altermagnets can be strongly energy dependent, a general effect that matters for applications: sharp bands at the Fermi level imply long carrier lifetimes favorable for spin transport. 

\begin{figure}[!htbp]
    \centering
    \includegraphics[width=1.0\linewidth]{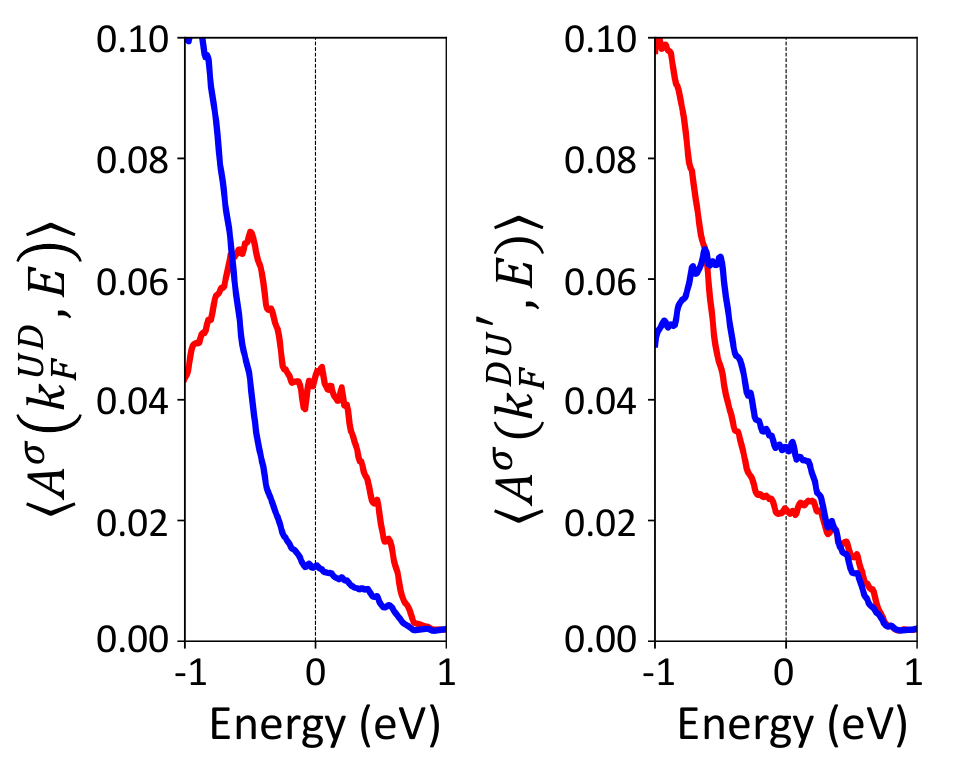}
\caption{Spin-resolved energy distribution curves $\langle A^{\sigma}(\mathbf{k},E)\rangle$ for 25\% Sb at the outermost Fermi momenta $\mathbf{k}=k_F^{UD}$ (left) and $\mathbf{k}=k_F^{DU'}$ (right) marked in Fig.~\ref{fig6:spectral}. Red and blue denote $\sigma=\uparrow$ and $\downarrow$, respectively.}
    \label{fig8:spinmix}
\end{figure}

Next, we turn to the more strongly doped case of 25\% Sb, where the altermagnetic character is still recognizable in Fig.~\ref{fig6:spectral}: the outermost band crossing the Fermi level carries predominantly spin-up weight at $k_F^{UD}$ and predominantly spin-down weight at $k_F^{DU'}$, but both are strongly broadened. This broadening degrades the spin selectivity, as quantified by the spin-resolved energy distribution curves in Fig.~\ref{fig8:spinmix}. At both momenta, the quasiparticle peak at $E_F$ appears as a clear shoulder of a second quasiparticle peak centered around $-0.5$~eV. At $k_F^{UD}$, the spin-up and spin-down spectral weights at $E_F$ are about 0.046 and 0.013~eV$^{-1}$, respectively, a minority-spin fraction of about 22\%. At $k_F^{DU'}$, the spin-down and spin-up spectral weights at $E_F$ are about 0.033 and 0.022~eV$^{-1}$, respectively, a minority-spin fraction of about 40\%. The asymmetry between the two momenta allows us to separate two contributions. Disorder broadening mixes the two spin channels equally at both momenta, whereas the partial ferromagnetic character of the ground state adds spin-up weight everywhere, reducing the minority fraction at $k_F^{UD}$ and increasing it at $k_F^{DU'}$. We write the minority fractions as $f_{\mathrm{broad}} \mp f_{\mathrm{FM}}$, where $f_{\mathrm{broad}}$ is the contribution of disorder broadening, assumed equal at both momenta, and $f_{\mathrm{FM}}$ the shift due to the partial ferromagnetic character. This gives $f_{\mathrm{broad}} \approx 31\%$ and $f_{\mathrm{FM}} \approx 9\%$, so disorder broadening is the dominant source of spin mixing.

Disorder broadening thus emerges as a new factor controlling spin selectivity, beyond the commonly considered spin splitting. Although 25\% Sb represents an extreme case close to the altermagnetic-to-ferromagnetic crossover, disorder-induced spin mixing will occur to a lesser extent at any finite doping. This has direct consequences for applications. In spintronic devices, readout through tunneling magnetoresistance depends steeply on the spin polarization at the Fermi level, so even moderate spin mixing can substantially reduce the signal. Disorder broadening also affects superconductivity. In a clean altermagnet, the spin splitting leaves a single spin channel at a given momentum on the Fermi surface, which blocks spin-singlet pairing between $\mathbf{k}$ and $-\mathbf{k}$ in favor of equal-spin triplet pairing. Disorder-induced spin mixing restores weight in the opposite spin channel at the same momentum, reopening the spin-singlet channel so that it can compete with triplet pairing. Disorder broadening is therefore an important consideration for both spintronic and unconventional superconducting applications of altermagnets.

\section{\label{sec:discussion} Discussion}

We have introduced a machine-learning framework that renormalizes first-principles calculations of strongly disordered magnets onto disordered spin-fermion models, and applied it to Sb-doped MnTe. By combining a GNN that learns explicit exchange couplings with a GNN that learns the electronic Hamiltonian in a given magnetic configuration, we obtained the magnetic ground states and momentum-resolved electronic structure of large disordered supercells across the altermagnetic-to-ferromagnetic crossover. Trained on small cells and benchmarked against DFT, the learned models predicted millions of inequivalent, long-range exchange couplings in many large disordered supercells, whose noncollinear ground states would be prohibitive to obtain with direct first-principles calculations. This revealed that altermagnetism persists up to about 25\% Sb, and that disorder broadening, alongside the spin splitting, controls spin selectivity, adding a new dimension to the design of altermagnets.

A natural next step is to include atomic relaxations, which we omitted here to keep the generation of training data tractable. Since both networks take atomic positions as input, this can be achieved by training on relaxed supercells and combining the framework with machine-learned force fields~\cite{Behler2007,Zhang2018DeePMD,Batzner2022}: structures would be relaxed first, after which GNN1 provides the exchange couplings and GNN2 the electronic Hamiltonian. Beyond relaxations, GNN1 can also be extended from isotropic Heisenberg exchange to the full exchange tensor, including Dzyaloshinskii--Moriya, Kitaev and $\Gamma$ interactions. This would enable simulations of moir\'e-engineered spin textures, such as the noncollinear moir\'e magnetism and skyrmions in twisted CrI$_3$ bilayers~\cite{Tong2018MoireSkyrmion,Akram2021MoireSkyrmion,Song2021MoireCrI3}. Such systems are typically modeled by taking the exchange at each point of the moir\'e cell from untwisted bilayers with the same local stacking~\cite{Bistritzer2011,kim2023ab}. Learning the exchange tensor directly from the local atomic environment would avoid this approximation, including at larger twist angles and with lattice relaxation. For the electronic model, training GNN2 on Wannier-downfolded Hamiltonians would further reduce the electronic Hilbert space, but generating such training data in practice requires automated Wannierization~\cite{Vitale2020Wannier,Qiao2023Projectability}.

Energy-based machine-learning models that also depend on the spin directions~\cite{Yu2024SpinGNN,Yuan2024MagNet,Yang2023DeePSPIN} offer a complementary route to the magnetic model, in which the magnetic interactions are learned from total energies of constrained spin configurations. This energy-based approach naturally captures interactions beyond bilinear exchange, such as biquadratic and multi-spin terms, as well as spin--lattice coupling, and provides the energies and forces needed for spin--lattice dynamics. The magnetic force theorem used here, in turn, extracts all bilinear couplings from a single calculation as a small perturbation around a reference state. This is advantageous for long-range exchange in metals, where an energy-based approach requires many constrained calculations per structure that may converge to different self-consistent solutions. It is also advantageous for systems with very weak exchange, such as $f$-electron magnets~\cite{PRB109_014425} or magnets with well-separated magnetic ions~\cite{Abdeldaim2024RPSO}. There, energy mapping must extract sub-meV couplings as small differences of large total energies, whereas the magnetic force theorem computes each coupling directly as a second-order response, so that even small couplings retain their relative accuracy. A systematic comparison of the two approaches, including their training efficiency, would clarify which is best suited to which class of materials.

Finally, the disordered spin-fermion model presented here is one realization of combining renormalization with machine-learned Hamiltonians, but the concept extends beyond spin-fermion models and beyond DFT as the high-level theory. For example, exact diagonalization of cluster models derived from first principles provides exchange tensors, whose derivatives with respect to atomic displacements yield spin--phonon couplings~\cite{Dhakal2024}. Learning these tensors as functions of atomic positions, analogous to electron--phonon couplings from machine-learned DFT Hamiltonians~\cite{Li2024DeepHDFPT}, could extend such couplings to disordered materials. This would be highly relevant for Kitaev spin liquid materials, where the half-quantized thermal Hall effect can only be observed in the presence of edge disorder~\cite{Ye2018ThermalHall,Zhang2025RuCl3}. As another example, Bethe--Salpeter calculations~\cite{Rohlfing2000BSE,Haber2023ExcitonWannier} could be renormalized onto effective models for moir\'e excitons in transition-metal dichalcogenide bilayers, which form tunable arrays of quantum emitters~\cite{Tran2019Moire,Seyler2019Moire,Jin2019Moire,Alexeev2019Moire}. More generally, the approach demonstrated here opens a broad route: any accurate but expensive first-principles method can, in principle, be renormalized onto an effective model of the degrees of freedom that matter. Machine learning then carries first-principles accuracy from small cells to the large, structurally complex systems relevant to experiments and devices.

\section{\label{sec:methods} Methods}

\subsection{First-principles calculations}

DFT calculations were performed with OpenMX~\cite{Ozaki2003}, using pseudoatomic orbital basis sets Mn: s2p2d1, Te: s3p2d2, Sb: s3p2d1, the Perdew-Burke-Ernzerhof (PBE) exchange-correlation functional \cite{perdew1996generalized} and a Hubbard $U = 3$~eV on Mn~d within the Dudarev scheme~\cite{Dudarev1998}, chosen consistently with our previous study, where it ensured dynamical stability and reasonable lattice parameters and phonons for MnTe and MnSb~\cite{Raghuvanshi2026}. Spin--orbit coupling was not included. $k$-meshes were generated with a density of 70 points per $2\pi \mathrm{\AA}^{-1}$ along each reciprocal lattice direction. We used an energy cutoff of 400~Ry. Training supercells contained randomly placed Sb atoms and had random lattice constants and angles, with cell angles $\alpha$ and $\beta$ drawn between $70^\circ$ and $110^\circ$ and $\gamma$ between $50^\circ$ and $130^\circ$, and cell volumes chosen to give about 80 atoms on average. For the electronic training data, random noncollinear spin configurations were stabilized with a penalty functional \cite{kurz2004ab} of 0.5 eV.

\subsection{Exchange couplings}

Exchange couplings were computed with the magnetic force theorem~\cite{Liechtenstein1987,Katsnelson2000} as implemented in the Jx code~\cite{yoon2020jx,yoon2018reliability}, using a smaller k-mesh with 2/3 density used in the self-consistent calculations to reduce computational cost. For each Sb concentration, the response was evaluated relative to the altermagnetic reference state. The couplings are defined for unit spins ($S=1$), so that they correspond to $JS^2$ in a model with spin length $S$.

\subsection{Graph neural networks}

The spin model (GNN1) builds on DeepH~\cite{Li2022DeepH}, which we modified to train on spin Hamiltonians, with exchange couplings from the magnetic force theorem as edge targets instead of electronic Hamiltonian blocks. The modified code, RenormHNet, is available at ~\cite{renormhnet} together with example input files. We also tested the E(3)-equivariant DeepH-E3~\cite{Gong2023DeepHE3}, but found it computationally more expensive. GNN1 was trained on 510 supercells: 10 at 2\% Sb and 50 at each Sb concentration from 5 to 50\% in steps of 5\%. These were randomly split into 90\% for training and 10\% for validation. The learning rate was initialized to $1.6 \times 10^{-3}$ and multiplied by 0.8 at epochs 40, 80, and 120, and by 0.6 at epoch 150. We used a cutoff radius of 10~\AA, covering couplings up to $J_{10}$. The validation loss reached its minimum after 57 epochs, at which point the root-mean-square error over all couplings in the validation set was 0.356~meV; beyond this point it increased while the training loss continued to decrease, indicating overfitting, and we therefore used the model at 57 epochs.

The electronic model (GNN2) uses xDeepH~\cite{Li2023xDeepH} without modification, with a cutoff radius of 8~\AA. GNN2 was trained on 2040 supercells: 4 chemical configurations at 2\% Sb and 20 at each Sb concentration from 5 to 50\% in steps of 5\%, each with 10 random spin configurations. These were randomly split into 90\% for training and 10\% for validation. The initial learning rate was set to $2 \times 10^{-3}$. Whenever the validation loss failed to improve for 20 consecutive epochs, the best checkpoint was restored and the learning rate was multiplied by 0.8. The model with the lowest validation loss, at epoch 264, was used for subsequent calculations, with a root-mean-square error of 2.88~meV for the Hamiltonian matrix elements in the validation set.

\subsection{Magnetic ground states}

Magnetic ground states were obtained by damped Landau--Lifshitz--Gilbert relaxation of classical unit spins~\cite{Gilbert2004}, using a phenomenological damping parameter $\lambda=1$ and a time step of $\Delta t=0.01$, until the maximum torque fell below $1\times10^{-10}$ meV. For each supercell, we started from 50 random spin configurations and kept the lowest-energy solution.

\subsection{Band unfolding}

Spectral functions were unfolded into the primitive Brillouin zone following Ref.~\cite{Ku2010unfolding},
\begin{equation}
A^{\sigma}(\mathbf{k},E)=\sum_{n}\sum_{\mathbf{K}J}\left|\langle \mathbf{k}n\sigma|\mathbf{K}J\rangle\right|^{2}\,\delta(\varepsilon_{\mathbf{K}J}-E),
\end{equation}
where $\varepsilon_{\mathbf{K}J}$ and $|\mathbf{K}J\rangle$ are the supercell eigenvalues and eigenstates, and $|\mathbf{k}n\sigma\rangle$ are primitive-cell Bloch states of orbital $n$ and spin $\sigma$. The overlaps are obtained from the coefficients of $|\mathbf{K}J\rangle$ in orthonormal supercell Bloch orbitals $|\mathbf{K}N\rangle$,
\begin{equation}
\begin{split}
\langle \mathbf{k}n\sigma|\mathbf{K}J\rangle={}&\frac{1}{\sqrt{N_p}}\sum_{N} e^{-i\mathbf{k}\cdot\mathbf{r}(N)}\,\delta_{n\sigma,\,n'(N)}\\
&\times\delta_{[\mathbf{k}-\mathbf{K}]}\,\langle \mathbf{K}N|\mathbf{K}J\rangle,
\end{split}
\end{equation}
where $\mathbf{r}(N)$ is the position of orbital $N$ in the supercell, $n'(N)$ its orbital and spin index in the primitive cell, $N_p$ the number of primitive cells per supercell, and $\delta_{[\mathbf{k}-\mathbf{K}]}$ selects the supercell wave vector $\mathbf{K}$ that unfolds to $\mathbf{k}$. The orbitals $|\mathbf{K}N\rangle$ must be orthonormal for the unfolded spectral functions to be normalized. Because the pseudoatomic orbitals $|\mathbf{K}\tilde{N}\rangle$ of OpenMX are non-orthogonal, we constructed L\"owdin-orthogonalized orbitals $|\mathbf{K}N\rangle=\sum_{M}|\mathbf{K}\tilde{M}\rangle\,[S(\mathbf{K})^{-1/2}]_{MN}$, with overlap matrix $S_{MN}(\mathbf{K})=\langle \mathbf{K}\tilde{M}|\mathbf{K}\tilde{N}\rangle$, in which the coefficients become $\langle \mathbf{K}N|\mathbf{K}J\rangle=\sum_{M}[S(\mathbf{K})^{-1/2}]_{NM}\,\langle \mathbf{K}\tilde{M}|\mathbf{K}J\rangle$. The $\delta$ function was replaced by a Lorentzian with a half width at half maximum of 20~meV.  The spectral functions along the $U$--$D$--$U'$ path were computed at 200 $\mathbf{k}$ points. The spectral functions were averaged over disorder configurations $c$,
\begin{equation}
\langle A^{\sigma}(\mathbf{k},E)\rangle=\frac{1}{N_c}\sum_{c=1}^{N_c}A^{\sigma}_{c}(\mathbf{k},E),
\end{equation}
with $N_c = 100$ non-orthogonal supercells with random lattice constants and angles, about 120 atoms on average, and random Sb configurations for each Sb concentration.

\section*{Data availability}
The following data are available on Zenodo~\cite{Kiem2026Zenodo}: the trained network weights of GNN1 and GNN2; all 510 supercell structures and their DFT exchange couplings used to train GNN1, together with selected OpenMX and Jx input and output files; and all 2040 supercell structures and spin configurations used to train GNN2, together with the OpenMX input files.

\section*{Code availability}
RenormHNet, our modified version of DeepH for learning exchange couplings, is available on GitHub~\cite{renormhnet}. The input files and hyperparameters for training, together with the full GNN1 training set of 510 supercells and their exchange couplings, are available on Zenodo~\cite{Kiem2026Zenodo}.

\section*{Author contributions}
D.H.K. and T.B. designed the study, analyzed and interpreted the data, and wrote the manuscript. D.H.K. developed the code, performed the calculations and curated the data. P.R.R. provided initial input and information for the calculations. All authors contributed to the discussions and reviewed the manuscript.

\begin{acknowledgments}
This work was supported by the U.S. Department of Energy, Office of Science, Basic Energy Sciences, Materials Sciences and Engineering Division. This research used resources of the Compute and Data Environment for Science (CADES) at the Oak Ridge National Laboratory, which is supported by the Office of Science of the U.S. Department of Energy under Contract No. DE-AC05-00OR22725. A portion of this research used computational resources at the Center for Nanophase Materials Sciences, which is a Department of Energy (DOE) Office of Science User Facility. Claude (Opus 5.5, Anthropic) was used, as directed by the authors, for text editing and literature search; all scientific content was verified by the authors.
\end{acknowledgments}

\bibliographystyle{apsrev4-2}
\bibliography{GNNMnTe_reference}

@article{Haber2023ExcitonWannier,
  author  = {Haber, Jonah B. and Qiu, Diana Y. and da Jornada, Felipe H. and Neaton, Jeffrey B.},
  title   = {Maximally Localized Exciton {W}annier Functions for Solids},
  journal = {Phys. Rev. B}, volume = {108}, pages = {125118}, year = {2023},
  doi     = {10.1103/PhysRevB.108.125118}
}

@article{Qian2026MACEH,
  author  = {Qian, Chen and Vitartas, Valdas and Kermode, James R. and Maurer, Reinhard J.},
  title   = {Equivariant Electronic Hamiltonian Prediction with Many-Body Message Passing},
  journal = {npj Comput. Mater.},
  volume  = {12},
  pages   = {169},
  year    = {2026},
  doi     = {10.1038/s41524-026-02020-1}
}

@misc{Kiem2026Zenodo,
  author       = {Kiem, Do Hoon and Raghuvanshi, Parul R. and Parker, David S. and Lindsay, Lucas and Cooper, Valentino R. and Berlijn, Tom},
  title        = {Data for ``Machine-Learning Spin-Fermion Models for Strongly Disordered Magnets: Sb-Doped MnTe''},
  howpublished = {Zenodo},
  year         = {2026},
  doi          = {10.5281/zenodo.24406792}
}

@article{Schutt2019SchNOrb,
  author  = {Sch{\"u}tt, K. T. and Gastegger, M. and Tkatchenko, A. and M{\"u}ller, K.-R. and Maurer, R. J.},
  title   = {Unifying Machine Learning and Quantum Chemistry with a Deep Neural Network for Molecular Wavefunctions},
  journal = {Nat. Commun.}, volume = {10}, pages = {5024}, year = {2019},
  doi     = {10.1038/s41467-019-12875-2}
}

@article{Naka2019,
  author  = {Naka, Makoto and Hayami, Satoru and Kusunose, Hiroaki and Yanagi, Yuki and Motome, Yukitoshi and Seo, Hitoshi},
  title   = {Spin Current Generation in Organic Antiferromagnets},
  journal = {Nat. Commun.},
  volume  = {10},
  pages   = {4305},
  year    = {2019},
  doi     = {10.1038/s41467-019-12229-y}
}

@article{Hayami2019,
  author  = {Hayami, Satoru and Yanagi, Yuki and Kusunose, Hiroaki},
  title   = {Momentum-Dependent Spin Splitting by Collinear Antiferromagnetic Ordering},
  journal = {J. Phys. Soc. Jpn.},
  volume  = {88},
  pages   = {123702},
  year    = {2019},
  doi     = {10.7566/JPSJ.88.123702}
}

@article{Hayami2020,
  author  = {Hayami, Satoru and Yanagi, Yuki and Kusunose, Hiroaki},
  title   = {Bottom-Up Design of Spin-Split and Reshaped Electronic Band Structures in Spin-Orbit-Coupling Free Antiferromagnets: Procedure on the Basis of Augmented Multipoles},
  journal = {Phys. Rev. B}, volume = {102}, pages = {144441}, year = {2020},
  doi     = {10.1103/PhysRevB.102.144441}
}

@article{Ozaki2003,
  author  = {Ozaki, T.},
  title   = {Variationally Optimized Atomic Orbitals for Large-Scale Electronic Structures},
  journal = {Phys. Rev. B},
  volume  = {67},
  pages   = {155108},
  year    = {2003},
  doi     = {10.1103/PhysRevB.67.155108}
}

@article{Dudarev1998,
  author  = {Dudarev, S. L. and Botton, G. A. and Savrasov, S. Y. and Humphreys, C. J. and Sutton, A. P.},
  title   = {Electron-Energy-Loss Spectra and the Structural Stability of Nickel Oxide: An {LSDA+U} Study},
  journal = {Phys. Rev. B},
  volume  = {57},
  pages   = {1505},
  year    = {1998},
  doi     = {10.1103/PhysRevB.57.1505}
}

@article{Liu2024MnTe,
  author  = {Liu, Zheyuan and Ozeki, Makoto and Asai, Shinichiro and Itoh, Shinichi and Masuda, Takatsugu},
  title   = {Chiral Split Magnon in Altermagnetic {MnTe}},
  journal = {Phys. Rev. Lett.},
  volume  = {133},
  pages   = {156702},
  year    = {2024},
  doi     = {10.1103/PhysRevLett.133.156702}
}

@article{Yamaguchi1992MnSb,
  author  = {Yamaguchi, Y. and Tomiyoshi, S. and Harada, M. and Shirane, G.},
  title   = {Magnons and Phonons in {MnSb}},
  journal = {J. Magn. Magn. Mater.},
  volume  = {103},
  pages   = {50--52},
  year    = {1992},
  doi     = {10.1016/0304-8853(92)90234-F}
}

@article{Mazin2023MnTe,
  author  = {Mazin, I. I.},
  title   = {Altermagnetism in {MnTe}: Origin, Predicted Manifestations, and Routes to Detwinning},
  journal = {Phys. Rev. B},
  volume  = {107},
  pages   = {L100418},
  year    = {2023},
  doi     = {10.1103/PhysRevB.107.L100418}
}

@article{Alaei2024MnTe,
  author  = {Alaei, Mojtaba and Sobieszczyk, Pawel and Ptok, Andrzej and Rezaei, Nafise and Oganov, Artem R. and Qaiumzadeh, Alireza},
  title   = {Origin of {$A$}-Type Antiferromagnetism and Chiral Split Magnons in Altermagnetic $\alpha$-{MnTe}},
  journal = {Phys. Rev. B},
  volume  = {111},
  pages   = {104416},
  year    = {2025},
  doi     = {10.1103/PhysRevB.111.104416}
}

@misc{Dhakal2024,
  author        = {Dhakal, Ramesh and Kaib, David A. S. and Choi, Kate and Biswas, Sananda and Valent\'i, Roser and Winter, Stephen M.},
  title         = {Theory of Intrinsic Phonon Thermal {H}all Effect in $\alpha$-{RuCl$_3$}},
  year          = {2024},
  eprint        = {2407.00660},
  archivePrefix = {arXiv},
  url           = {https://arxiv.org/abs/2407.00660}
}

@article{Abdeldaim2024RPSO,
  author  = {Abdeldaim, Aly H. and Gretarsson, Hlynur and Day, Sarah J. and Le, M. Duc and Stenning, Gavin B. G. and Manuel, Pascal and Perry, Robin S. and Tsirlin, Alexander A. and Nilsen, G{\o}ran J. and Clark, Lucy},
  title   = {Kitaev Interactions through Extended Superexchange Pathways in the $j_{\mathrm{eff}}=1/2$ {Ru$^{3+}$} Honeycomb Magnet {RuP$_3$SiO$_{11}$}},
  journal = {Nat. Commun.},
  volume  = {15},
  pages   = {9778},
  year    = {2024},
  doi     = {10.1038/s41467-024-53900-3}
}

@article{PRB109_014425,
  author  = {Scheie, A. O. and Kamiya, Y. and Zhang, Hao and Lee, Sangyun and Woods, A. J. and Ajeesh, M. O. and Gonzalez, M. G. and Bernu, B. and Villanova, J. W. and Xing, J. and Huang, Q. and Zhang, Qingming and Ma, Jie and Choi, Eun Sang and Pajerowski, D. M. and Zhou, Haidong and Sefat, A. S. and Okamoto, S. and Berlijn, T. and Messio, L. and Movshovich, R. and Batista, C. D. and Tennant, D. A.},
  title   = {Nonlinear Magnons and Exchange {H}amiltonians of the Delafossite Proximate Quantum Spin Liquid Candidates {KYbSe$_2$} and {NaYbSe$_2$}},
  journal = {Phys. Rev. B},
  volume  = {109},
  pages   = {014425},
  year    = {2024},
  doi     = {10.1103/PhysRevB.109.014425}
}

@article{Tong2018MoireSkyrmion,
  author  = {Tong, Qingjun and Liu, Fei and Xiao, Jiang and Yao, Wang},
  title   = {Skyrmions in the Moir\'e of van der {W}aals {2D} Magnets},
  journal = {Nano Lett.},
  volume  = {18},
  pages   = {7194--7199},
  year    = {2018},
  doi     = {10.1021/acs.nanolett.8b03315}
}

@article{Akram2021MoireSkyrmion,
  author  = {Akram, Muhammad and LaBollita, Harrison and Dey, Dibyendu and Kapeghian, Jesse and Erten, Onur and Botana, Antia S.},
  title   = {Moir\'e Skyrmions and Chiral Magnetic Phases in Twisted {CrX$_3$} ({X} = {I}, {Br}, and {Cl}) Bilayers},
  journal = {Nano Lett.},
  volume  = {21},
  pages   = {6633--6639},
  year    = {2021},
  doi     = {10.1021/acs.nanolett.1c02096}
}

@article{Song2021MoireCrI3,
  author  = {Song, Tiancheng and Sun, Qi-Chao and Anderson, Eric and Wang, Chong and Qian, Jimin and Taniguchi, Takashi and Watanabe, Kenji and McGuire, Michael A. and St{\"o}hr, Rainer and Xiao, Di and Cao, Ting and Wrachtrup, J{\"o}rg and Xu, Xiaodong},
  title   = {Direct Visualization of Magnetic Domains and Moir\'e Magnetism in Twisted {2D} Magnets},
  journal = {Science},
  volume  = {374},
  pages   = {1140--1144},
  year    = {2021},
  doi     = {10.1126/science.abj7478}
}

@article{Bistritzer2011,
  author  = {Bistritzer, Rafi and MacDonald, Allan H.},
  title   = {Moir\'e Bands in Twisted Double-Layer Graphene},
  journal = {Proc. Natl. Acad. Sci. U.S.A.},
  volume  = {108},
  pages   = {12233--12237},
  year    = {2011},
  doi     = {10.1073/pnas.1108174108}
}

@article{Vitale2020Wannier,
  author  = {Vitale, Valerio and Pizzi, Giovanni and Marrazzo, Antimo and Yates, Jonathan R. and Marzari, Nicola and Mostofi, Arash A.},
  title   = {Automated High-Throughput {W}annierisation},
  journal = {npj Comput. Mater.},
  volume  = {6},
  pages   = {66},
  year    = {2020},
  doi     = {10.1038/s41524-020-0312-y}
}

@article{Qiao2023Projectability,
  author  = {Qiao, Junfeng and Pizzi, Giovanni and Marzari, Nicola},
  title   = {Projectability Disentanglement for Accurate and Automated Electronic-Structure {H}amiltonians},
  journal = {npj Comput. Mater.},
  volume  = {9},
  pages   = {208},
  year    = {2023},
  doi     = {10.1038/s41524-023-01146-w}
}

@article{Li2024DeepHDFPT,
  author  = {Li, He and Tang, Zechen and Fu, Jingheng and Dong, Wen-Han and Zou, Nianlong and Gong, Xiaoxun and Duan, Wenhui and Xu, Yong},
  title   = {Deep-Learning Density Functional Perturbation Theory},
  journal = {Phys. Rev. Lett.},
  volume  = {132},
  pages   = {096401},
  year    = {2024},
  doi     = {10.1103/PhysRevLett.132.096401}
}

@article{Ye2018ThermalHall,
  author  = {Ye, Mengxing and Hal\'asz, G\'abor B. and Savary, Lucile and Balents, Leon},
  title   = {Quantization of the Thermal {H}all Conductivity at Small {H}all Angles},
  journal = {Phys. Rev. Lett.},
  volume  = {121},
  pages   = {147201},
  year    = {2018},
  doi     = {10.1103/PhysRevLett.121.147201}
}

@article{Zhang2025RuCl3,
  author        = {Zhang, Heda and Hal\'asz, G\'abor B. and Ghosh, Sujoy and Jesse, Stephen and Ward, Thomas Z. and Tennant, David A. and McGuire, Michael A. and Yan, Jiaqiang},
  title         = {Evidence of Chiral Fermion Edge Modes through Geometric Engineering of Thermal {H}all Effect in $\alpha$-{RuCl$_3$}},
  journal       = {Phys. Rev. Lett.},
  volume        = {136},
  pages         = {226301},
  year          = {2026},
  doi           = {10.1103/rn48-7j6y}
}

@article{Rohlfing2000BSE,
  author  = {Rohlfing, Michael and Louie, Steven G.},
  title   = {Electron-Hole Excitations and Optical Spectra from First Principles},
  journal = {Phys. Rev. B},
  volume  = {62},
  pages   = {4927},
  year    = {2000},
  doi     = {10.1103/PhysRevB.62.4927}
}

@article{Tran2019Moire,
  author  = {Tran, Kha and others},
  title   = {Evidence for Moir\'e Excitons in van der {W}aals Heterostructures},
  journal = {Nature},
  volume  = {567},
  pages   = {71--75},
  year    = {2019},
  doi     = {10.1038/s41586-019-0975-z}
}

@article{Seyler2019Moire,
  author  = {Seyler, Kyle L. and Rivera, Pasqual and Yu, Hongyi and Wilson, Nathan P. and Ray, Essance L. and Mandrus, David G. and Yan, Jiaqiang and Yao, Wang and Xu, Xiaodong},
  title   = {Signatures of Moir\'e-Trapped Valley Excitons in {MoSe$_2$/WSe$_2$} Heterobilayers},
  journal = {Nature},
  volume  = {567},
  pages   = {66--70},
  year    = {2019},
  doi     = {10.1038/s41586-019-0957-1}
}

@article{Jin2019Moire,
  author  = {Jin, Chenhao and Regan, Emma C. and Yan, Aiming and Iqbal Bakti Utama, M. and Wang, Danqing and Zhao, Sihan and Qin, Ying and Yang, Sijie and Zheng, Zhiren and Shi, Shenyang and Watanabe, Kenji and Taniguchi, Takashi and Tongay, Sefaattin and Zettl, Alex and Wang, Feng},
  title   = {Observation of Moir\'e Excitons in {WSe$_2$/WS$_2$} Heterostructure Superlattices},
  journal = {Nature},
  volume  = {567},
  pages   = {76--80},
  year    = {2019},
  doi     = {10.1038/s41586-019-0976-y}
}

@article{Alexeev2019Moire,
  author  = {Alexeev, Evgeny M. and Ruiz-Tijerina, David A. and Danovich, Mark and Hamer, Matthew J. and Terry, Daniel J. and Nayak, Pramoda K. and Ahn, Seongjoon and Pak, Sangyeon and Lee, Juwon and Sohn, Jung Inn and Molas, Maciej R. and Koperski, Maciej and Watanabe, Kenji and Taniguchi, Takashi and Novoselov, Kostya S. and Gorbachev, Roman V. and Shin, Hyeon Suk and Fal'ko, Vladimir I. and Tartakovskii, Alexander I.},
  title   = {Resonantly Hybridized Excitons in Moir\'e Superlattices in van der {W}aals Heterostructures},
  journal = {Nature},
  volume  = {567},
  pages   = {81--86},
  year    = {2019},
  doi     = {10.1038/s41586-019-0986-9}
}

@article{Berlijn2011NaxCoO2,
  author  = {Berlijn, Tom and Volja, Dmitri and Ku, Wei},
  title   = {Can Disorder Alone Destroy the $e_g'$ Hole Pockets of {Na$_x$CoO$_2$}? A {W}annier Function Based First-Principles Method for Disordered Systems},
  journal = {Phys. Rev. Lett.},
  volume  = {106},
  pages   = {077005},
  year    = {2011},
  doi     = {10.1103/PhysRevLett.106.077005}
}

@article{Gilbert2004,
  author  = {Gilbert, T. L.},
  title   = {A Phenomenological Theory of Damping in Ferromagnetic Materials},
  journal = {IEEE Trans. Magn.},
  volume  = {40},
  pages   = {3443},
  year    = {2004},
  doi     = {10.1109/TMAG.2004.836740}
}

@article{Szuszkiewicz2006,
  author  = {Szuszkiewicz, W. and Dynowska, E. and Witkowska, B. and Hennion, B.},
  title   = {Spin-Wave Measurements on Hexagonal {MnTe} of {NiAs}-Type Structure by Inelastic Neutron Scattering},
  journal = {Phys. Rev. B},
  volume  = {73},
  pages   = {104403},
  year    = {2006},
  doi     = {10.1103/PhysRevB.73.104403}
}

@article{Liechtenstein1987,
  author  = {Liechtenstein, A. I. and Katsnelson, M. I. and Antropov, V. P. and Gubanov, V. A.},
  title   = {Local Spin Density Functional Approach to the Theory of Exchange Interactions in Ferromagnetic Metals and Alloys},
  journal = {J. Magn. Magn. Mater.},
  volume  = {67},
  pages   = {65--74},
  year    = {1987},
  doi     = {10.1016/0304-8853(87)90721-9}
}

@article{Katsnelson2000,
  author  = {Katsnelson, M. I. and Lichtenstein, A. I.},
  title   = {First-Principles Calculations of Magnetic Interactions in Correlated Systems},
  journal = {Phys. Rev. B},
  volume  = {61},
  pages   = {8906},
  year    = {2000},
  doi     = {10.1103/PhysRevB.61.8906}
}

@article{Szilva2023,
  author  = {Szilva, Attila and Kvashnin, Yaroslav and Stepanov, Evgeny A. and Nordstr{\"o}m, Lars and Eriksson, Olle and Lichtenstein, Alexander I. and Katsnelson, Mikhail I.},
  title   = {Quantitative Theory of Magnetic Interactions in Solids},
  journal = {Rev. Mod. Phys.},
  volume  = {95},
  pages   = {035004},
  year    = {2023},
  doi     = {10.1103/RevModPhys.95.035004}
}

@misc{SI,
  key  = {SI},
  note = {See Supplementary Information for additional benchmarks of the exchange interactions $J_4$ through $J_9$, the benchmark of the electronic model at 50\% Sb, and the Brillouin zone with the $k$-path used for the spectral functions}
}

@article{Wang2021TBML,
  author  = {Wang, Zifeng and Ye, Shizhuo and Wang, Hao and He, Jin and Huang, Qijun and Chang, Sheng},
  title   = {Machine Learning Method for Tight-Binding {H}amiltonian Parameterization from Ab-Initio Band Structure},
  journal = {npj Comput. Mater.},
  volume  = {7},
  pages   = {11},
  year    = {2021},
  doi     = {10.1038/s41524-020-00490-5}
}

@article{Schattauer2022,
  author  = {Schattauer, Christoph and Todorovi\'c, Milica and Ghosh, Kunal and Rinke, Patrick and Libisch, Florian},
  title   = {Machine Learning Sparse Tight-Binding Parameters for Defects},
  journal = {npj Comput. Mater.},
  volume  = {8},
  pages   = {116},
  year    = {2022},
  doi     = {10.1038/s41524-022-00791-x}
}

@article{Cignoni2024,
  author  = {Cignoni, Edoardo and Suman, Divya and Nigam, Jigyasa and Cupellini, Lorenzo and Mennucci, Benedetta and Ceriotti, Michele},
  title   = {Electronic Excited States from Physically Constrained Machine Learning},
  journal = {ACS Cent. Sci.},
  volume  = {10},
  pages   = {637--648},
  year    = {2024},
  doi     = {10.1021/acscentsci.3c01480}
}

@article{Choudhary2025SlaKoNet,
  author  = {Choudhary, Kamal},
  title   = {{SlaKoNet}: A Unified {S}later--{K}oster Tight-Binding Framework Using Neural Network Infrastructure for the Periodic Table},
  journal = {J. Phys. Chem. Lett.},
  volume  = {16},
  pages   = {11109--11119},
  year    = {2025},
  doi     = {10.1021/acs.jpclett.5c02456}
}

@misc{Zhong2023HamGNNpp,
  author        = {Zhong, Yang and Zhang, Binhua and Yu, Hongyu and Gong, Xingao and Xiang, Hongjun},
  title         = {Accelerating the Electronic-Structure Calculation of Magnetic Systems by Equivariant Neural Networks},
  year          = {2023},
  eprint        = {2306.01558},
  archivePrefix = {arXiv},
  primaryClass  = {cond-mat.mtrl-sci},
  url           = {https://arxiv.org/abs/2306.01558}
}

@article{Yu2024SpinGNN,
  author  = {Yu, Hongyu and Zhong, Yang and Hong, Liangliang and Xu, Changsong and Ren, Wei and Gong, Xingao and Xiang, Hongjun},
  title   = {Spin-Dependent Graph Neural Network Potential for Magnetic Materials},
  journal = {Phys. Rev. B},
  volume  = {109},
  pages   = {144426},
  year    = {2024},
  doi     = {10.1103/PhysRevB.109.144426}
}

@article{Yuan2024MagNet,
  author        = {Yuan, Zilong and Xu, Zhiming and Li, He and Cheng, Xinle and Tao, Honggeng and Tang, Zechen and Zhou, Zhiyuan and Duan, Wenhui and Xu, Yong},
  title         = {Equivariant Neural Network Force Fields for Magnetic Materials},
  journal       = {Quantum Front.},
  volume        = {3},
  pages         = {8},
  year          = {2024},
  doi           = {10.1007/s44214-024-00055-3}
}

@article{Yang2023DeePSPIN,
  author  = {Yang, Teng and Cai, Zefeng and Huang, Zhengtao and Tang, Wenlong and Shi, Ruosong and Godfrey, Andy and Liu, Hanxing and Lin, Yuanhua and Nan, Ce-Wen and Ye, Meng and Zhang, Linfeng and Wang, Ke and Wang, Han and Xu, Ben},
  title   = {Deep Learning Illuminates Spin and Lattice Interaction in Magnetic Materials},
  journal = {Phys. Rev. B},
  volume  = {110},
  pages   = {064427},
  year    = {2024},
  doi     = {10.1103/PhysRevB.110.064427}
}

@article{Ruderman1954,
  author  = {Ruderman, M. A. and Kittel, C.},
  title   = {Indirect Exchange Coupling of Nuclear Magnetic Moments by Conduction Electrons},
  journal = {Phys. Rev.},
  volume  = {96},
  pages   = {99},
  year    = {1954},
  doi     = {10.1103/PhysRev.96.99}
}

@article{Kasuya1956,
  author  = {Kasuya, Tadao},
  title   = {A Theory of Metallic Ferro- and Antiferromagnetism on {Z}ener's Model},
  journal = {Prog. Theor. Phys.},
  volume  = {16},
  pages   = {45},
  year    = {1956},
  doi     = {10.1143/PTP.16.45}
}

@article{Yosida1957,
  author  = {Yosida, Kei},
  title   = {Magnetic Properties of {Cu-Mn} Alloys},
  journal = {Phys. Rev.},
  volume  = {106},
  pages   = {893},
  year    = {1957},
  doi     = {10.1103/PhysRev.106.893}
}

@article{Ku2010unfolding,
  author  = {Ku, Wei and Berlijn, Tom and Lee, Chi-Cheng},
  title   = {Unfolding First-Principles Band Structures},
  journal = {Phys. Rev. Lett.},
  volume  = {104},
  pages   = {216401},
  year    = {2010},
  doi     = {10.1103/PhysRevLett.104.216401}
}

@article{Popescu2010,
  author  = {Popescu, Voicu and Zunger, Alex},
  title   = {Effective Band Structure of Random Alloys},
  journal = {Phys. Rev. Lett.},
  volume  = {104},
  pages   = {236403},
  year    = {2010},
  doi     = {10.1103/PhysRevLett.104.236403}
}

@article{Raghuvanshi2026,
  author  = {Raghuvanshi, P. R. and Li, X. and Berlijn, T. and Ghosh, A. and Parker, D. S. and Cooper, V. R. and Lindsay, L.},
  title   = {Disorder-Induced Magnetoelastic Behavior of {MnTe$_x$Sb$_y$Bi$_{1-x-y}$} Alloys},
  journal = {npj Comput. Mater.},
  year    = {2026},
  doi     = {10.1038/s41524-026-02206-7}
}

@article{Mu2019MnTe,
  author  = {Mu, Sai and Hermann, Rapha{\"e}l P. and Gorsse, St{\'e}phane and Zhao, Huaizhou and Manley, Michael E. and Fishman, Randy S. and Lindsay, Lucas},
  title   = {Phonons, Magnons, and Lattice Thermal Transport in Antiferromagnetic Semiconductor {MnTe}},
  journal = {Phys. Rev. Mater.},
  volume  = {3},
  pages   = {025403},
  year    = {2019},
  doi     = {10.1103/PhysRevMaterials.3.025403}
}

@article{Krempasky2024,
  author  = {Krempask\'y, J. and \v{S}mejkal, L. and D'Souza, S. W. and Hajlaoui, M. and Springholz, G. and Uhl\'i\v{r}ov\'a, K. and Alarab, F. and Constantinou, P. C. and Strocov, V. and Usanov, D. and Pudelko, W. R. and Gonz\'alez-Hern\'andez, R. and Birk Hellenes, A. and Jansa, Z. and Reichlov\'a, H. and \v{S}ob\'a\v{n}, Z. and Gonzalez Betancourt, R. D. and Wadley, P. and Sinova, J. and Kriegner, D. and Min\'ar, J. and Dil, J. H. and Jungwirth, T.},
  title   = {Altermagnetic Lifting of {K}ramers Spin Degeneracy},
  journal = {Nature},
  volume  = {626},
  pages   = {517--522},
  year    = {2024},
  doi     = {10.1038/s41586-023-06907-7}
}

@article{Lee2024MnTe,
  author  = {Lee, Suyoung and Lee, Sangjae and Jung, Saegyeol and Jung, Jiwon and Kim, Donghan and Lee, Yeonjae and Seok, Byeongjun and Kim, Jaeyoung and Park, Byeong Gyu and \v{S}mejkal, Libor and Kang, Chang-Jong and Kim, Changyoung},
  title   = {Broken {K}ramers Degeneracy in Altermagnetic {MnTe}},
  journal = {Phys. Rev. Lett.},
  volume  = {132},
  pages   = {036702},
  year    = {2024},
  doi     = {10.1103/PhysRevLett.132.036702}
}

@article{Continenza2001,
  author  = {Continenza, A. and Picozzi, S. and Geng, W. T. and Freeman, A. J.},
  title   = {Coordination and Chemical Effects on the Structural, Electronic, and Magnetic Properties in {Mn} Pnictides},
  journal = {Phys. Rev. B},
  volume  = {64},
  pages   = {085204},
  year    = {2001},
  doi     = {10.1103/PhysRevB.64.085204}
}

@article{Taylor2015MnSb,
  author  = {Taylor, A. E. and Berlijn, T. and Hahn, S. E. and May, A. F. and Williams, T. J. and Poudel, L. and Calder, S. and Fishman, R. S. and Stone, M. B. and Aczel, A. A. and Cao, H. B. and Lumsden, M. D. and Christianson, A. D.},
  title   = {Influence of Interstitial {Mn} on Magnetism in the Room-Temperature Ferromagnet {Mn$_{1+\delta}$Sb}},
  journal = {Phys. Rev. B},
  volume  = {91},
  pages   = {224418},
  year    = {2015},
  doi     = {10.1103/PhysRevB.91.224418}
}

@article{Marenkin2018,
  author  = {Marenkin, S. F. and Kochura, A. V. and Izotov, A. D. and Vasil'ev, M. G.},
  title   = {Manganese Pnictides {MnP}, {MnAs}, and {MnSb} Are Ferromagnetic Semimetals: Preparation, Structure, and Properties (a Survey)},
  journal = {Russ. J. Inorg. Chem.},
  volume  = {63},
  pages   = {1753--1763},
  year    = {2018},
  doi     = {10.1134/S0036023618140036}
}

@article{Wan2025metallicAM,
  author  = {Wan, Xuhao and Mandal, Subhasish and Guo, Yuzheng and Haule, Kristjan},
  title   = {High-Throughput Search for Metallic Altermagnets by Embedded Dynamical Mean Field Theory},
  journal = {Phys. Rev. Lett.},
  volume  = {135},
  pages   = {106501},
  year    = {2025},
  doi     = {10.1103/k47t-23gp}
}

@article{Sufyan2026,
  author  = {Sufyan, Ali and Marfoua, Brahim and Larsson, J. Andreas and van Loon, Erik and Armiento, Rickard},
  title   = {High-Throughput Quantification of Altermagnetic Band Splitting},
  journal = {Phys. Rev. Mater.},
  volume  = {10},
  pages   = {044407},
  year    = {2026},
  doi     = {10.1103/mmdm-hrj4}
}

@inproceedings{Gilmer2017,
  author    = {Gilmer, Justin and Schoenholz, Samuel S. and Riley, Patrick F. and Vinyals, Oriol and Dahl, George E.},
  title     = {Neural Message Passing for Quantum Chemistry},
  booktitle = {Proceedings of the 34th International Conference on Machine Learning},
  series    = {Proceedings of Machine Learning Research},
  volume    = {70},
  pages     = {1263--1272},
  year      = {2017},
  url       = {http://proceedings.mlr.press/v70/gilmer17a.html}
}

@article{Behler2007,
  author  = {Behler, J{\"o}rg and Parrinello, Michele},
  title   = {Generalized Neural-Network Representation of High-Dimensional Potential-Energy Surfaces},
  journal = {Phys. Rev. Lett.},
  volume  = {98},
  pages   = {146401},
  year    = {2007},
  doi     = {10.1103/PhysRevLett.98.146401}
}

@article{Zhang2018DeePMD,
  author  = {Zhang, Linfeng and Han, Jiequn and Wang, Han and Car, Roberto and E, Weinan},
  title   = {Deep Potential Molecular Dynamics: A Scalable Model with the Accuracy of Quantum Mechanics},
  journal = {Phys. Rev. Lett.},
  volume  = {120},
  pages   = {143001},
  year    = {2018},
  doi     = {10.1103/PhysRevLett.120.143001}
}

@article{Batzner2022,
  author  = {Batzner, Simon and Musaelian, Albert and Sun, Lixin and Geiger, Mario and Mailoa, Jonathan P. and Kornbluth, Mordechai and Molinari, Nicola and Smidt, Tess E. and Kozinsky, Boris},
  title   = {E(3)-Equivariant Graph Neural Networks for Data-Efficient and Accurate Interatomic Potentials},
  journal = {Nat. Commun.},
  volume  = {13},
  pages   = {2453},
  year    = {2022},
  doi     = {10.1038/s41467-022-29939-5}
}

@article{Li2022DeepH,
  author  = {Li, He and Wang, Zun and Zou, Nianlong and Ye, Meng and Xu, Runzhang and Gong, Xiaoxun and Duan, Wenhui and Xu, Yong},
  title   = {Deep-Learning Density Functional Theory {H}amiltonian for Efficient Ab Initio Electronic-Structure Calculation},
  journal = {Nat. Comput. Sci.},
  volume  = {2},
  pages   = {367--377},
  year    = {2022},
  doi     = {10.1038/s43588-022-00265-6}
}

@article{Gong2023DeepHE3,
  author  = {Gong, Xiaoxun and Li, He and Zou, Nianlong and Xu, Runzhang and Duan, Wenhui and Xu, Yong},
  title   = {General Framework for E(3)-Equivariant Neural Network Representation of Density Functional Theory {H}amiltonian},
  journal = {Nat. Commun.},
  volume  = {14},
  pages   = {2848},
  year    = {2023},
  doi     = {10.1038/s41467-023-38468-8}
}

@article{Zhong2023HamGNN,
  author  = {Zhong, Yang and Yu, Hongyu and Su, Mao and Gong, Xingao and Xiang, Hongjun},
  title   = {Transferable Equivariant Graph Neural Networks for the {H}amiltonians of Molecules and Solids},
  journal = {npj Comput. Mater.},
  volume  = {9},
  pages   = {182},
  year    = {2023},
  doi     = {10.1038/s41524-023-01130-4}
}

@inproceedings{Yu2023QHNet,
  author    = {Yu, Haiyang and Xu, Zhao and Qian, Xiaofeng and Qian, Xiaoning and Ji, Shuiwang},
  title     = {Efficient and Equivariant Graph Networks for Predicting Quantum {H}amiltonian},
  booktitle = {Proceedings of the 40th International Conference on Machine Learning},
  series    = {Proceedings of Machine Learning Research},
  volume    = {202},
  pages     = {40412--40424},
  year      = {2023},
  url       = {https://proceedings.mlr.press/v202/yu23i.html}
}

@article{Gu2024DeePTB,
  author  = {Gu, Qiangqiang and Zhouyin, Zhanghao and Pandey, Shishir Kumar and Zhang, Peng and Zhang, Linfeng and E, Weinan},
  title   = {Deep Learning Tight-Binding Approach for Large-Scale Electronic Simulations at Finite Temperatures with Ab Initio Accuracy},
  journal = {Nat. Commun.},
  volume  = {15},
  pages   = {6772},
  year    = {2024},
  doi     = {10.1038/s41467-024-51006-4}
}

@article{Li2023xDeepH,
  author  = {Li, He and Tang, Zechen and Gong, Xiaoxun and Zou, Nianlong and Duan, Wenhui and Xu, Yong},
  title   = {Deep-Learning Electronic-Structure Calculation of Magnetic Superstructures},
  journal = {Nat. Comput. Sci.},
  volume  = {3},
  pages   = {321--327},
  year    = {2023},
  doi     = {10.1038/s43588-023-00424-3}
}

@article{Smejkal2022landscape,
  author  = {\v{S}mejkal, Libor and Sinova, Jairo and Jungwirth, Tomas},
  title   = {Emerging Research Landscape of Altermagnetism},
  journal = {Phys. Rev. X},
  volume  = {12},
  pages   = {040501},
  year    = {2022},
  doi     = {10.1103/PhysRevX.12.040501}
}

@article{Smejkal2022beyond,
  author  = {\v{S}mejkal, Libor and Sinova, Jairo and Jungwirth, Tomas},
  title   = {Beyond Conventional Ferromagnetism and Antiferromagnetism: A Phase with Nonrelativistic Spin and Crystal Rotation Symmetry},
  journal = {Phys. Rev. X},
  volume  = {12},
  pages   = {031042},
  year    = {2022},
  doi     = {10.1103/PhysRevX.12.031042}
}

@article{GonzalezHernandez2021,
  author  = {Gonz\'alez-Hern\'andez, Rafael and \v{S}mejkal, Libor and V\'yborn\'y, Karel and Yahagi, Yuta and Sinova, Jairo and Jungwirth, Tom\'a\v{s} and \v{Z}elezn\'y, Jakub},
  title   = {Efficient Electrical Spin Splitter Based on Nonrelativistic Collinear Antiferromagnetism},
  journal = {Phys. Rev. Lett.},
  volume  = {126},
  pages   = {127701},
  year    = {2021},
  doi     = {10.1103/PhysRevLett.126.127701}
}

@article{Smejkal2022GMR,
  author  = {\v{S}mejkal, Libor and Hellenes, Anna Birk and Gonz\'alez-Hern\'andez, Rafael and Sinova, Jairo and Jungwirth, Tomas},
  title   = {Giant and Tunneling Magnetoresistance in Unconventional Collinear Antiferromagnets with Nonrelativistic Spin-Momentum Coupling},
  journal = {Phys. Rev. X},
  volume  = {12},
  pages   = {011028},
  year    = {2022},
  doi     = {10.1103/PhysRevX.12.011028}
}

@article{Bai2022SST,
  author  = {Bai, H. and Han, L. and Feng, X. Y. and Zhou, Y. J. and Su, R. X. and Wang, Q. and Liao, L. Y. and Zhu, W. X. and Chen, X. Z. and Pan, F. and Fan, X. L. and Song, C.},
  title   = {Observation of Spin Splitting Torque in a Collinear Antiferromagnet {RuO$_2$}},
  journal = {Phys. Rev. Lett.},
  volume  = {128},
  pages   = {197202},
  year    = {2022},
  doi     = {10.1103/PhysRevLett.128.197202}
}

@article{Karube2022,
  author  = {Karube, Shutaro and Tanaka, Takahiro and Sugawara, Daichi and Kadoguchi, Naohiro and Kohda, Makoto and Nitta, Junsaku},
  title   = {Observation of Spin-Splitter Torque in Collinear Antiferromagnetic {RuO$_2$}},
  journal = {Phys. Rev. Lett.},
  volume  = {129},
  pages   = {137201},
  year    = {2022},
  doi     = {10.1103/PhysRevLett.129.137201}
}

@article{Zhu2023TSC,
  author  = {Zhu, Di and Zhuang, Zheng-Yang and Wu, Zhigang and Yan, Zhongbo},
  title   = {Topological Superconductivity in Two-Dimensional Altermagnetic Metals},
  journal = {Phys. Rev. B},
  volume  = {108},
  pages   = {184505},
  year    = {2023},
  doi     = {10.1103/PhysRevB.108.184505}
}

@article{Brekke2023,
  author  = {Brekke, Bj{\o}rnulf and Brataas, Arne and Sudb{\o}, Asle},
  title   = {Two-Dimensional Altermagnets: Superconductivity in a Minimal Microscopic Model},
  journal = {Phys. Rev. B},
  volume  = {108},
  pages   = {224421},
  year    = {2023},
  doi     = {10.1103/PhysRevB.108.224421}
}

@article{Ghorashi2024,
  author  = {Ghorashi, Sayed Ali Akbar and Hughes, Taylor L. and Cano, Jennifer},
  title   = {Altermagnetic Routes to Majorana Modes in Zero Net Magnetization},
  journal = {Phys. Rev. Lett.},
  volume  = {133},
  pages   = {106601},
  year    = {2024},
  doi     = {10.1103/PhysRevLett.133.106601}
}

@article{yoon2020jx,
  title={{Jx}: An open-source software for calculating magnetic interactions based on magnetic force theory},
  author={Yoon, Hongkee and Kim, Taek Jung and Sim, Jae-Hoon and Han, Myung Joon},
  journal={Comput. Phys. Commun.},
  volume={247},
  pages={106927},
  year={2020},
  doi={10.1016/j.cpc.2019.106927}
}

@article{yoon2018reliability,
  title={Reliability and applicability of magnetic-force linear response theory: Numerical parameters, predictability, and orbital resolution},
  author={Yoon, Hongkee and Kim, Taek Jung and Sim, Jae-Hoon and Jang, Seung Woo and Ozaki, Taisuke and Han, Myung Joon},
  journal={Phys. Rev. B},
  volume={97},
  number={12},
  pages={125132},
  year={2018},
  doi={10.1103/PhysRevB.97.125132}
}

@article{kurz2004ab,
  title={Ab initio treatment of noncollinear magnets with the full-potential linearized augmented plane wave method},
  author={Kurz, Ph and F{\"o}rster, F and Nordstr{\"o}m, Lars and Bihlmayer, Gustav and Bl{\"u}gel, Stefan},
  journal={Phys. Rev. B},
  volume={69},
  number={2},
  pages={024415},
  year={2004},
  doi={10.1103/PhysRevB.69.024415}
}

@article{perdew1996generalized,
  title={Generalized gradient approximation made simple},
  author={Perdew, John P and Burke, Kieron and Ernzerhof, Matthias},
  journal={Phys. Rev. Lett.},
  volume={77},
  number={18},
  pages={3865},
  year={1996},
  doi={10.1103/PhysRevLett.77.3865}
}

@article{kim2023ab,
  title={Ab initio spin hamiltonian and topological noncentrosymmetric magnetism in twisted bilayer {CrI$_3$}},
  author={Kim, Kyoung-Min and Kiem, Do Hoon and Bednik, Grigory and Han, Myung Joon and Park, Moon Jip},
  journal={Nano Lett.},
  volume={23},
  number={13},
  pages={6088--6094},
  year={2023},
  doi={10.1021/acs.nanolett.3c01529}
}

@misc{renormhnet,
  howpublished = {\url{https://github.com/DHKiem/RenormHNet.git}},
}

\end{document}